\documentclass[11pt,a4paper]{article}
\usepackage{jheppub}
\usepackage{natbib}
\usepackage{slashed}

\usepackage{multirow}
\usepackage{verbatim}

\title{Octet baryon masses in next-to-next-to-next-to-leading order
covariant baryon chiral perturbation theory}

\author[a]{X.-L. Ren,}
\author[a]{L. S. Geng,}
\author[b]{J. Martin Camalich}
\author[a,c,d]{J. Meng}
\author[e]{and H. Toki}
\affiliation[a]{Research Center for Nuclear Science and Technology \& School of Physics and
Nuclear Energy Engineering, Beihang University,
Beijing 100191, China}
\affiliation[b]{Department of Physics and Astronomy, University of Sussex,
BN1 9QH, Brighton, UK}
\affiliation[c]{State Key Laboratory of Nuclear Physics and Technology, School of Physics,
Peking University,\\
Beijing 100871, China}
\affiliation[d]{Department of Physics, University of Stellenbosch,
 Stellenbosch, South Africa}
 \affiliation[e]{Research Center for Nuclear Physics (RCNP), Osaka University, Ibaraki, Osaka 567-0047, Japan}

\emailAdd{lisheng.geng@buaa.edu.cn}

\abstract{
  We study the ground-state octet baryon masses and sigma terms using the covariant baryon chiral perturbation theory (ChPT) with
  the extended-on-mass-shell (EOMS) renormalization scheme up to next-to-next-to-next-to-leading order (N$^3$LO). By adjusting the available $19$ low-energy constants (LECs), a reasonable fit of
  the $n_f=2+1$ lattice quantum chromodynamics (LQCD) results from the PACS-CS, LHPC, HSC, QCDSF-UKQCD and NPLQCD collaborations is achieved. Finite-volume corrections to the lattice data are calculated self-consistently. Our study shows that the N$^3$LO BChPT describes better the light quark mass evolution of the lattice data than the NNLO BChPT does and the various lattice simulations seem to be
   consistent with each other. We also predict the pion and strangeness sigma terms of the octet baryons using the LECs determined in the fit of
  their masses. The predicted pion-  and strangeness-nucleon sigma terms are $\sigma_{\pi N}=43(1)(6)$ MeV and  $\sigma_{s N}=126(24)(54)$ MeV,  respectively.}


\begin{document}
\maketitle
\flushbottom

\section{Introduction}

Studies of baryon spectroscopy play an important role in understanding the nonperturbative nature of quantum chromodynamics (QCD).
In the past few years, Lattice QCD simulations~\cite{Wilson1974_Phys.Rev.D10_2445, Gattringer2010__} have made remarkable progress in studies of non-perturbative
strong-interaction physics. Recently, the lowest-lying baryon spectrum, composed of up, down and strange quarks, has been studied by various LQCD collaborations~\cite{Alexandrouitetal.2009_Phys.Rev.D80_114503, Durritetal.2009_Science322_1224, Aokiitetal.(PACS-CSCollaboration)2009_Phys.Rev.D79_034503, Aokiitetal.(PACS-CSCollaboration)2010_Phys.Rev.D81_074503, Walker-Louditelal.2009_Phys.Rev.D79_054502, Linitetal.(HSCCollaboration)2009_Phys.Rev.D79_034502, Bietenholzitetal.2010_Phys.Lett.B690_436--441, Bietenholzitetal.(QCDSF-UKQCDCollaboration)2011_PhysicalReviewD84_054509, Beaneitetal.(NPLQCDCollaboration)2011_PhysicalReviewD84_014507} . However, because these calculations adopt different lattice setup and all of them lead to the same continuum theory, it is crucial to test whether the results for baryon masses are consistent with each other~\cite{Beringeritetal.(ParticleDataGroup)2012_Phys.Rev.D86_010001}. On the other hand, since lattice QCD simulations are performed in a finite hypercube and  with larger than physical light quarks masses~\cite{Fodor2012_Rev.Mod.Phys.84_449}, the final results can only be obtained by extrapolating to the physical point (chiral extrapolation) and infinite space-time (finite volume corrections). Chiral perturbation theory (ChPT) provides a useful framework to perform such extrapolations and to study the induced uncertainties.

ChPT, as the low energy effective field theory of QCD, is based on effective Lagrangian techniques with an expansion in powers of external momenta and light quark masses, constrained by chiral symmetry and its breaking pattern~\cite{Weinberg1979_PhysicaA96_327--340,Gasser1984_Ann.Phys.(N.Y.)158_142--210,
Gasser1985_Nucl.Phys.B250_465--516,Gasser1988_Nucl.Phys.B307_779--853,Leutwyler1994_Lecture_,
Bernard1995_Int.J.Mod.Phys.E4_193--346,Pich1995_Rep.Prog.Phys.58_57,
Ecker1995_Prog.Part.Nucl.Phys.35_84,Pich1998_arXiv_, Bernard2007_Annu.Rev.Nucl.Part.Sci.57_33--60, Bernard2008_Prog.Part.Nucl.Phys.60_82--160,Scherer2012__}. Its applications in the mesonic sector have been
rather successful at least in the two flavor sector of $u$ and $d$ quarks.
But the extension to the one-baryon sector  turns out to be non-trivial. Because baryon masses do not vanish in the chiral limit, a systematic power counting (PC) is absent~\cite{Gasser1988_Nucl.Phys.B307_779--853}. In order to restore the chiral power counting, the so-called Heavy-Baryon (HB) ChPT  was first proposed by Jenkins and Manohar~\cite{Jenkins1991_Phys.Lett.B255_558--562}, considering baryons as heavy static sources. Although this approach provides a strict power-counting, the heavy baryon expansion is non-relativistic,
which might lead to pathologies in certain cases and is found to converge rather slowly in the three flavor sector of $u$, $d$, and $s$ quarks. Later, covariant BChPT implementing a consistent PC with different renormalization methods have been developed, such as the infrared (IR)~\cite{Becher1999_Eur.Phys.J.C9_643--671} and the extended-on-mass-shell (EOMS) \cite{Gegelia1999_Phys.Rev.D60_114038,Fuchs2003_Phys.Rev.D68_056005} renormalization schemes.  In addition to the afore-mentioned dimensional renormalization schemes (MS and its derivatives), to speed up the convergence of BChPT, other renormalization/regularization schemes are also proposed, e.g., the cutoff scheme~\cite{Bernard2004_Nucl.Phys.A732_149--170}, the finite range regulator (FRR) method~\cite{Young2003_Prog.Part.Nucl.Phys.50_399--417, Leinweber2003_Phys.Rev.Lett.92_24,Young2010_Phys.Rev.D81_014503},  and the partial summation approach \cite{Semke2006_Nucl.Phys.A778_153--180}.

In the past decades, the ground-state (g.s.) octet baryon masses have been studied extensively~\cite{Jenkins1992_Nucl.Phys.B368_190,Bernard1993_Z.PhysikC60_111--119, Banerjee1995_Phys.Rev.D52_11, Borasoy1996_Ann.Phys.(N.Y.)254_192--232,Walker-Loud2004_Nucl.Phys.A747_476--507, Ellis1999_Phys.Rev.C61_015205, Frink2004_JHEP07_028, Frink2005_Eur.Phys.J.A.24_395, Lehnhart2004_J.Phys.G:Nucl.Part.Phys.31_89, MartinCamalich2010_Phys.Rev.D82_074504, Young2010_Phys.Rev.D81_014503, Semke2006_Nucl.Phys.A778_153--180,Semke2007_Nucl.Phys.A789_251--259, Semke2012_Phys.Rev.D85_034001,Bruns:2012eh,Lutz2012_Phys.Rev.D86_091502(R)}.
It is found that SU(3) HBChPT converges rather slowly ~\cite{Ishikawa2009_Phys.Rev.D80_054502}. Furthermore, most calculations are performed only up to NNLO because of the many unknown low-energy constants (LECs) at N$^3$LO except those of Refs.~\cite{Borasoy1996_Ann.Phys.(N.Y.)254_192--232,Walker-Loud2004_Nucl.Phys.A747_476--507, Frink2004_JHEP07_028, Semke2012_Phys.Rev.D85_034001,Bruns:2012eh,Lutz2012_Phys.Rev.D86_091502(R)}. Regarding chiral extrapolations, Young and Thomas~\cite{Young2010_Phys.Rev.D81_014503} obtained very good
 results using the FRR scheme up to NNLO by fitting the LHPC \cite{Walker-Louditelal.2009_Phys.Rev.D79_054502} and PACS-CS \cite{Aokiitetal.(PACS-CSCollaboration)2009_Phys.Rev.D79_034503} lattice data. In Ref. \cite{Ishikawa2009_Phys.Rev.D80_054502}, we applied the NNLO EOMS-BChPT to analyze the same lattice data and found that the EOMS-BChPT can provide a better description of lattice data and is more suitable for chiral extrapolation purposes than HBChPT and NLO BChPT. Recently, using a partial summation scheme up to N$^3$LO, Semke and Lutz~\cite{Semke2012_Phys.Rev.D85_034001,Scherer2012__,Lutz2012_Phys.Rev.D86_091502(R)} found that the BMW~\cite{Durritetal.2009_Science322_1224}, HSC~\cite{Linitetal.(HSCCollaboration)2009_Phys.Rev.D79_034502}, PACS-CS~\cite{Aokiitetal.(PACS-CSCollaboration)2009_Phys.Rev.D79_034503}, LHPC~\cite{Walker-Louditelal.2009_Phys.Rev.D79_054502}, and QCDSF-UKQCD~\cite{Bietenholzitetal.(QCDSF-UKQCDCollaboration)2011_PhysicalReviewD84_054509} lattice results can be well described.

On the other hand, up to now, a simultaneous description of all the $n_f=2+1$ lattice data with finite-volume effects taken into account self-consistently is still missing.~\footnote{ In Ref.~\cite{Lutz2012_Phys.Rev.D86_091502(R)}, Semke and Lutz showed that their partial summation approach can reproduce
the results of the HSC and QCDSF-UKQCD collaborations by fitting the BMW, PACS and LHP data.}  Such a study is necessary for a clarification of the convergence problem and for testing the consistency between different lattice simulations. Furthermore, it also provides a good opportunity to determine/constrain the many unknown LECs of BChPT at N$^3$LO.

In this work we study the g.s. octet baryon masses and sigma terms using the EOMS-BChPT up to N$^3$LO. Finite-volume corrections to the lattice data are calculated self-consistently and are found to be important in order that a good fit of the lattice data can be achieved. Unlike Refs.~\cite{Young2010_Phys.Rev.D81_014503,Semke2012_Phys.Rev.D85_034001}, the contributions of virtual decuplet baryons will not be explicitly included, because their effects can not be disentangled from those of virtual octet baryons due to the large number of unknown LECs and
because they are only expected to play a secondary role in determining the properties of octet baryons.\footnote{ However, their inclusion might be important for the determination of baryon sigma terms, as having been stressed in Ref.~\cite{Alarcon:2012nr} and will be the subject of a forthcoming work.} In order to fix all the $19$ LECs and test the consistency of current lattice calculations, we perform a simultaneous fit of all the publicly available $n_f=2+1$ LQCD data from the PACS-CS~\cite{Aokiitetal.(PACS-CSCollaboration)2009_Phys.Rev.D79_034503}, LHPC~\cite{Walker-Louditelal.2009_Phys.Rev.D79_054502}, HSC~\cite{Linitetal.(HSCCollaboration)2009_Phys.Rev.D79_034502}, QCDSF-UKQCD~\cite{Bietenholzitetal.(QCDSF-UKQCDCollaboration)2011_PhysicalReviewD84_054509} and NPLQCD~\cite{Beaneitetal.(NPLQCDCollaboration)2011_PhysicalReviewD84_014507} collaborations.

The paper is organized as follows. In Sec.~\ref{sec2} we collect the relevant terms of the chiral effective Lagrangians up to N$^3$LO, which involve $19$ unknown LECs. The explicit results of the g.s. octet baryon masses are provided in Sec.~\ref{sec3}, where the EOMS renormalization scheme is also briefly explained. Sec. \ref{sec4} focuses on the determination of the unknown LECs by fitting all the LQCD data and the chiral extrapolation of the octet baryon masses. We calculate the pion- and strangeness-baryon sigma terms in Sec.~\ref{sec5}.  A brief summary is  given in Sec.~\ref{sec6}.

\section{Power counting and effective Lagrangians}\label{sec2}
\subsection{Power counting}
In ChPT, the power counting (PC) provides a systematic organization of the effective Lagrangians and the corresponding loop diagrams within a perturbative expansion  in powers of $(p/\Lambda_{\rm \chi SB})^{n_{\rm ChPT}}$, where $p$ is a small momentum or scale and $\Lambda_{\rm ChPT}$ the chiral symmetry breaking scale. In the one-baryon sector, the chiral order, $n_{\rm ChPT}$, of a diagram with $L$ loops is calculated as
\begin{equation}\label{corder}
  n_{\rm ChPT} = 4L - 2N_{\phi} - N_{B} +\sum\limits_{k} kV_k,
\end{equation}
where $N_{\phi}(N_{B})$ is the number of internal meson (baryon) propagators, and $V_k$ is the number of vertices from $k$th-order Lagrangians. However, as mentioned in the introduction, in the covariant BChPT this systematic power counting is lost. That is to say, in the calculation of a loop diagram one may find
analytical terms with a chiral order lower than that determined by Eq.~(\ref{corder})\cite{Gasser1985_Nucl.Phys.B250_465--516}. In order to recover the power counting, we adopt the so called EOMS renormalization scheme. In this scheme, the lower-order power counting breaking pieces of the loop results are systematically absorbed into the available counter-terms~\cite{Becher1999_Eur.Phys.J.C9_643--671,Fuchs2003_Phys.Rev.D68_056005}. A detailed discussion of the EOMS renormalization scheme will be presented in Sec.~\ref{EOMS}.

\subsection{Chiral Lagrangians}
In this subsection, we collect the relevant chiral Lagrangians for the calculation of octet baryon masses in the three-flavor sector of
 $u$, $d$ and $s$ quarks up to N$^3$LO. The Lagrangians can be written as the sum of a mesonic part and a meson-baryon part:
\begin{equation}
  \mathcal{L}_{\mathrm{eff}}=\mathcal{L}_{\phi}^{(2)}+\mathcal{L}_\phi^{(4)}+\mathcal{L}_{\phi B}^{(1)}+\mathcal{L}_{\phi B}^{(2)} +\mathcal{L}_{\phi B}^{(3)}+\mathcal{L}_{\phi B}^{(4)},
\end{equation}
where the subscript $(i)$ denotes the corresponding chiral order $\mathcal{O}(p^i)$, $\phi=(\pi,~K,~\eta)$ represent the pseudoscalar Nambu-Goldstone boson fields, $B=(N,~\Lambda,~\Sigma,~\Xi)$ the g.s. octet baryons.

\subsubsection{Meson Lagrangians}
The lowest-order meson Lagrangian is given by
\begin{equation}\label{ml2}
  \mathcal{L}_{\phi}^{(2)}=\frac{F_{\phi}^2}{4}\langle D_{\mu}U(D^{\mu}U)^{\dag}\rangle
  +\frac{F_{\phi}^2}{4}\langle\chi U^{\dag}+U\chi^{\dag}\rangle,
\end{equation}
where $F_{\phi}$ is the pseudoscalar decay constant in the chiral  limit,
$\langle X\rangle$ stands for the trace in flavor space,
 $\chi=2 B_0\mathcal{M}$ accounts for explicit chiral symmetry breaking with $B_0=-\langle0|\bar{q}q|0\rangle/F_{\phi}^2$, $\mathcal{M}={\rm diag}(m_l,~m_l,~m_s)$, where we assumed perfect isospin symmetry, $m_u=m_d=m_l$. The $3\times 3$ unimodular, unitary matrix $U$ collects the pseudoscalar fields
\begin{equation}
  U(\phi)=u^2(\phi)={\rm exp}\left(i\frac{\phi}{F_{\phi}}\right),
\end{equation}
with
\begin{equation}
  \phi=\sum\limits_{a=1}^{8}\phi_a\lambda_a=\sqrt{2}\left(\begin{array}{ccc}
                       \frac{1}{\sqrt{2}}\pi^0+\frac{1}{\sqrt{6}}\eta & \pi^{+} & K^{+} \\
                       \pi^- & -\frac{1}{\sqrt{2}}\pi^0+\frac{1}{\sqrt{6}}\eta & K^0 \\
                       K^- & \bar{K}^0 & -\frac{2}{\sqrt{6}}\eta
                     \end{array}
  \right).
\end{equation}
Under SU(3)$_L$$\times$SU(3)$_R$, $U(x)$ transforms as $U\rightarrow U^{\prime}=LUR^{\dag}$, with $L, R\in$ SU(3)$_{L,R}$.

The most general meson Lagrangian at $\mathcal{O}(q^4)$ has the following form~\cite{Gasser1985_Nucl.Phys.B250_465--516}:
\begin{eqnarray}
  \mathcal{L}_{\phi}^{(4)}&=&L_1[\langle D_{\mu}U(D^{\mu}U)^{\dag}\rangle]^2
  +L_2\langle D_{\mu}U(D_{\nu}U)^{\dag} \rangle \langle D^{\mu}U(D^{\nu}U)^{\dag}\rangle\nonumber\\
  \qquad &&+L_3\langle D_{\mu}U(D^{\mu}U)^{\dag}D_{\nu}D(D^{\nu}U)^{\dag} \rangle
  +L_4\langle D_{\mu}U(D^{\mu}U)^{\dag}\rangle \langle \chi U^{\dag}+U\chi^{\dag}\rangle\nonumber\\
  \qquad &&+L_5\langle D_{\mu}U(D^{\mu}U)^{\dag}(\chi U^{\dag}+U\chi^{\dag})\rangle
  +L_6[\langle \chi U^{\dag}+U\chi^{\dag}\rangle]^2\nonumber\\
  \qquad &&+L_7[\langle \chi U^{\dag}-U\chi^{\dag}\rangle]^2 +L_8\langle U\chi^{\dag}U\chi^{\dag}+\chi U^{\dag}\chi U^{\dag}\rangle \nonumber\\
  \qquad &&-iL_9\langle f_{\mu\nu}^{R}D^{\mu}U(D^{\nu}U)^{\dag}+ f_{\mu\nu}^L(D^{\mu}U)^{\dag}D^{\nu}U\rangle
  +L_{10}\langle U f_{\mu\nu}^LU^{\dag}f_{R}^{\mu\nu}\rangle\nonumber\\
  \qquad &&+H_1\langle f_{\mu\nu}^Rf_{R}^{\mu\nu}+f_{\mu\nu}^Lf_L^{\mu\nu}\rangle +H_2\langle \chi\chi^{\dag}\rangle,
\end{eqnarray}
where $f_R^{\mu\nu}=\partial^\mu r^\nu-\partial^\nu r^\mu-i[r^\mu,r^\nu]$ and $f_L^{\mu\nu}=\partial^\mu l^\nu-\partial^\nu l^\mu-i[l^\mu,l^\nu]$ with
$r_{\mu}=v_{\mu}+a_{\mu}$, $l_{\mu}=v_{\mu}-a_{\mu}$ with $v_{\mu}$ and $a_{\mu}$ the external vector and axial currents. The LECs $L_i$'s are scale-dependent and absorb the infinities generated by the one-loop graphs.

\subsubsection{Meson-Baryon Lagrangians}
The effective meson-baryon Lagrangians contain terms of even and odd chiral orders,
\begin{equation}
  \mathcal{L}_{\phi B}^{\mathrm{eff}}=\mathcal{L}_{\phi B}^{(1)}+\mathcal{L}_{\phi B}^{(2)} +\mathcal{L}_{\phi B}^{(3)}+\mathcal{L}_{\phi B}^{(4)}.
\end{equation}
The lowest-order meson-baryon Lagrangian is
\begin{equation}\label{mbl1}
    \mathcal{L}_{\phi\mathrm{B}}^{(1)}=\langle\bar{B}(i\slashed D-m_0)B\rangle+\frac{D/F}{2}
    \langle\bar{B}\gamma^{\mu}\gamma_5[u_{\mu},B]_{\pm}\rangle,
\end{equation}
where $m_0$ denotes the baryon mass in the chiral limit, and the constants $D$ and $F$ are the axial-vector coupling constants, which are determined from the baryon semi-leptonic decays.  The traceless $3\times 3$ matrix $B$ contains the lowest-lying octet baryon fields
\begin{equation}
  B=\sum\limits_{a=1}^8\frac{B_a\lambda_a}{\sqrt{2}}=\left(\begin{array}{ccc}
            \frac{1}{\sqrt{2}}\Sigma^0+\frac{1}{\sqrt{6}}\Lambda & \Sigma^+ & p \\
            \Sigma^- & -\frac{1}{\sqrt{2}}\Sigma^0+\frac{1}{\sqrt{6}}\Lambda & n \\
            \Xi^- & \Xi^0 & -\frac{2}{\sqrt{6}}\Lambda
          \end{array}
  \right).
\end{equation}
Under SU(3)$_L\times$SU(3)$_R$, $B$ transforms as any matter field, $B\rightarrow B^{\prime}=KBK^{\dag}$,
with $K(U,L,R)$ the compensator field representing an element of the conserved subgroup SU(3)$_V$.
In equation (\ref{mbl1}), the covariant derivative of the baryon field is defined
\begin{equation}
  D_{\mu}B=\partial_{\mu}B+[\Gamma_{\mu},~B],
\end{equation}
\begin{equation}
  \Gamma_{\mu}=\frac{1}{2}\left\{u^{\dag}(\partial_{\mu}-i r_{\mu})+u(\partial_{\mu}-i l_{\mu})u^{\dag}\right\},
\end{equation}
and $u_{\mu}$ the axial current defined as
\begin{equation}
  u_{\mu}=i\left\{u^{\dag}(\partial_{\mu}-i r_{\mu})u-u(\partial_{\mu}-i l_{\mu})u^{\dag}\right\},
\end{equation}
where $u=\sqrt{U}$.

The meson-baryon Lagrangian at order $\mathcal{O}(p^2)$ can be written as
\begin{equation}
  \mathcal{L}_{\phi B}^{(2)}=\mathcal{L}_{\phi B}^{(2,~{\rm sb})}+{\mathcal{L}^{(2)}_{
  \phi B}}^{\prime}.
\end{equation}
This separation is motivated by the fact that the first part appears in the tree and loop graphs, whereas the latter only contributes via loops. The explicit chiral symmetry breaking part reads as
\begin{equation}
 \mathcal{L}_{\phi B}^{(2, {\rm sb})}=b_0 \langle \chi_+ \rangle \langle\bar{B}B\rangle
  +b_{D/F}\langle\bar{B}[\chi_+, B]_{\pm}\rangle,
\end{equation}
where $b_{0}$, $b_D$, and $b_F$ are LECs, and $\chi_{+}=u^{\dag}\chi u^{\dag}+u\chi^{\dag}u$. For the latter part, we take the same form as in Ref.~\cite{Oller2006_JHEP09_079}:
\begin{eqnarray}
  {\mathcal{L}^{(2)}_{\phi B}}^{\prime}&=& b_1\langle\bar{B}[u_{\mu},[u^{\mu},B]]\rangle + b_2\langle\bar{B}\{u_{\mu},\{u^{\mu}, B\}\}\rangle\nonumber\\
  \qquad &&+b_3\langle\bar{B}\{u_{\mu},[u^{\mu}, B]\}\rangle+b_4\langle\bar{B}B\rangle\langle u^{\mu}u_{\mu}\rangle\nonumber\\
  \qquad &&+ib_5\left(\langle\bar{B}[u^{\mu},[u^{\nu},\gamma_{\mu}D_{\nu}B]]\rangle-\langle\bar{B}
  \overleftarrow{D}_{\nu}[u^{\nu},[u^{\mu},\gamma_{\mu}B]]\right)\nonumber\\
  \qquad &&+ib_6\left(\langle\bar{B}[u^{\mu},\{u^{\nu},\gamma_{\mu}D_{\nu}B\}]\rangle-\langle\bar{B}
  \overleftarrow{D}_{\nu}\{u^{\nu},[u^{\mu},\gamma_{\mu}B]\}\right)\nonumber\\
  \qquad &&+ib_7\left(\langle\bar{B}\{u^{\mu},\{u^{\nu},\gamma_{\mu}D_{\nu}B\}\}\rangle-\langle\bar{B}
  \overleftarrow{D}_{\nu}\{u^{\nu},\{u^{\mu},\gamma_{\mu}B\}\}\rangle\right)\nonumber\\
  \qquad &&+ib_8\left(\langle\bar{B}\gamma_{\mu}D_{\nu}B\rangle-\langle\bar{B}
  \overleftarrow{D}_{\nu}\gamma_{\mu}B\rangle\right)\langle u^{\mu}u^{\nu}\rangle+\cdots,
\end{eqnarray}
 where $b_{1,\cdots,4}$ have dimension mass$^{-1}$ and $b_{5,\cdots,8}$ have dimension mass$^{-2}$.
If one works for a set of fixed quark masses (e.g., Ref.~\cite{Frink2005_Eur.Phys.J.A.24_395}), all terms with one or two covariant derivatives can be absorbed in the structures proportional to $b_{1,\cdots,4}$.
However, for our purposes, we need to retain all the terms because they lead to different quark mass dependencies.

The third chiral order Lagrangian  does not contribute to the baryon masses and the fourth-order effective Lagrangian relevant to our study is~\cite{Borasoy1996_Ann.Phys.(N.Y.)254_192--232}:
\begin{eqnarray}
  \mathcal{L}_{\phi B}^{(4)}  &=& d_1\langle\bar{B}[\chi_+, [\chi_+, B]]\rangle+d_2\langle\bar{B}[\chi_+, \{\chi_+, B\}]\rangle\nonumber\\
  \qquad &&+d_3\langle\bar{B}\{\chi_+,\{\chi_+, B\}\}\rangle+d_4\langle\bar{B}\chi_+\rangle\langle\chi_+B\rangle\nonumber\\
  \qquad &&+d_5\langle\bar{B}[\chi_+, B]\rangle \langle\chi_+\rangle+d_7\langle\bar{B}B\rangle\langle\chi_+\rangle^2\nonumber\\
  \qquad &&+d_8\langle\bar{B}B\rangle\langle\chi_+^2\rangle.
\end{eqnarray}

In total, up to N$^3$LO, we have 19 LECs: $m_0$,~$b_{0}$,~$b_D$,~$b_F$,~$b_{1-8}$ and $d_{1-5,~7,~8}$.

\section{Octet baryon masses}\label{sec3}

In this section, we evaluate the octet baryon masses up to N$^3$LO using the covariant BChPT, supplemented by the  EOMS renormalization scheme.

\subsection{Baryon self-energy up to N$^3$LO}
\begin{figure}[t]
\centering
\includegraphics[width=15cm]{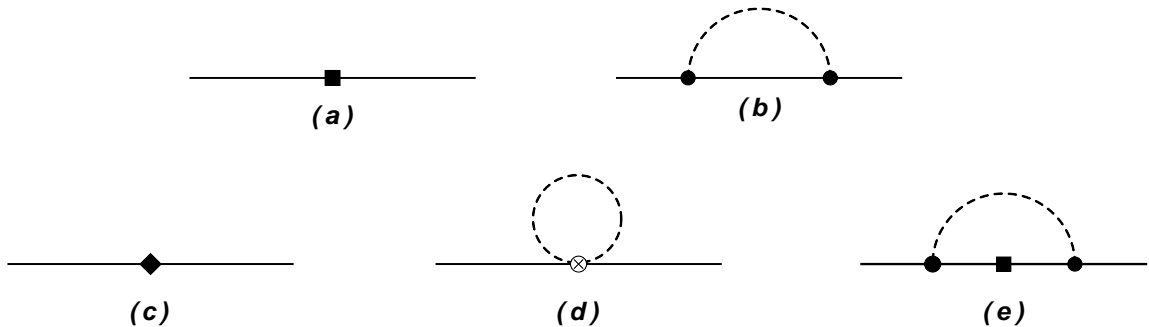}
\caption{Feynman diagrams contributing to the octet-baryon masses up to $\mathcal{O}(p^4)$ in the EOMS-BChPT. The solid lines correspond to octet-baryons and dashed lines refer to Goldstone bosons. The black boxes (diamonds) indicate second (fourth) order couplings. The solid dot (circle-cross) indicates an insertion from the dimension one (two) meson-baryon Lagrangians. Wave function renormalization diagrams are not explicitly shown but included in the calculation.}
\label{feydia}
\end{figure}

The physical baryon mass is defined at the baryon pole, $\slashed p=m_B$, in
the two-point function of the baryon field $\psi_B(x)$
\begin{equation}
  S_0(x)=-i\langle0|T[\psi_B(x)\bar{\psi}_B(0)]|0\rangle=\frac{1}{\slashed p-m_0-\Sigma(\slashed p)},
\end{equation}
where $m_0$ is the baryon mass in the chiral limit and $\Sigma(\slashed p)$ corresponds to the baryon self-energy
\begin{equation}
  m_B-m_0-\Sigma(\slashed p=m_B)=0, \quad \Rightarrow \quad m_B=m_0+\Sigma(\slashed p=m_B).
\end{equation}
The leading contribution to the self-energy, $\Sigma_a=m_B^{(2)}$, is of order $\mathcal{O}(p^2)$ [Fig.~\ref{feydia}(a)].
The self-energy $\Sigma_b=m_B^{(3)}$ of the one-loop diagram [Fig.~\ref{feydia}(b)] is of order $\mathcal{O}(p^3)$.
One tree diagram contribution from $\mathcal{L}_{\phi B}^{(4)}$ [Fig.~\ref{feydia}(c)] and
two loop diagrams [Figs.~\ref{feydia}(d,e)] are of order $\mathcal{O}(p^4)$, $m_B^{(4)}=\Sigma_c+\Sigma_d+\Sigma_e$. We remark that due to parity conservation, there are no first order contributions. The baryon mass up to fourth order in the chiral expansion can be expressed
\begin{equation}\label{chiexpand}
  m_B=m_0+m_B^{(2)}+m_B^{(3)}+m_B^{(4)}.
\end{equation}

The tree diagrams Fig.~\ref{feydia}(a,c) can be calculated straightforwardly. The corresponding results are shown below.
The three one-loop diagrams Figs.~\ref{feydia}(b), ~\ref{feydia}(d) and \ref{feydia}(e)  yield, generically,
\begin{eqnarray}
G_b &=& i~\int~\frac{d^4k}{(2\pi)^4}~\slashed k\gamma_5\frac{1}{\slashed p-\slashed k-m_0+i\epsilon}\slashed k \gamma_5\frac{1}{k^2-M_{\phi}^2+i\epsilon},\label{integral1} \\
G_d &=& i~\int~\frac{d^4k}{(2\pi)^4}~\{1,~k_{\mu}k^{\mu},~k_{\mu}k_{\nu}p^{\mu}\gamma^{\nu}\} \frac{1}{k^2-M_{\phi}^2+i\epsilon},\\
G_e &=& i~m_{B}^{(2)}\int~\frac{d^4k}{(2\pi)^4}~\slashed k\gamma_5\left(\frac{1}{\slashed p-\slashed k-m_0+i\epsilon}\right)^2\slashed k \gamma_5\frac{1}{k^2-M_{\phi}^2+i\epsilon},
\end{eqnarray}
where $M_{\phi}$ represents the mass of a  Nambu-Goldstone boson. Here, we want to mention that the integral of Eq.~\eqref{integral1} has been calculated in
Ref.~\cite{MartinCamalich2010_Phys.Rev.D82_074504}, and the results can be found there.  The above loop functions contain power-counting breaking (PCB) terms and therefore additional steps need to be taken to conserve a proper chiral power-counting scheme. Among the different approaches, the EOMS scheme has been shown to be superior to heavy-baryon or infrared approaches for a number of observables (at least at the one-loop level)~\cite{Geng2008_Phys.Rev.Lett.101_222002, Geng2011_Phys.Lett.B696_390, MartinCamalich2010_Phys.Rev.D82_074504, Geng2010_ChinesePhys.C34_1307}.

\subsection{The EOMS renormalization scheme}\label{EOMS}
The essence of the EOMS scheme is to perform an additional subtraction of PCB pieces beyond the $\widetilde{\rm MS}$ or $\overline{\rm MS}$ renormalization scheme. That is possible because the PCB terms appearing in a loop calculation are analytical and can always be removed by redefining the corresponding LECs. This is equivalent to removing the finite PCB pieces directly from the loop results. For an explicit calculation, the PC can be restored in two slightly different ways: (1) one can first perform the loop calculation analytically, and then remove the PCB terms, or (2) one can first perform an expansion in terms of the inverse heavy-baryon mass, $1/m_B$, calculate the PCB terms, and then subtract them from the full results. It should be noted that the first approach may not always work because one may not
be able to calculate the loop diagrams analytically, but since the PCB terms are finite and analytical, the second prescription should always work.

In the present study, we have explicitly checked that all the PCB terms appearing in our loop calculation can be removed by redefining the LECs introduced in the previous subsection.

\subsection{The mass formulas}
After calculating all the Feynman diagrams shown in Fig.~\ref{feydia} and subtracting the PCB terms using the EOMS renormalization scheme,
we obtain the full expressions of the octet baryon masses up to N$^3$LO.

At $\mathcal{O}(p^2)$ the tree level contribution provides the leading-order (LO) SU(3)-breaking corrections to the chiral limit octet baryon mass
\begin{equation}\label{mass2}
  m_B^{(2)}=\sum\limits_{\phi=\pi,~K}\xi_{B,\phi}^{(a)}M_{\phi}^2,
\end{equation}
where the coefficients $\xi_{B,\phi}^{(a)}$ are listed in Table~\ref{tree2coef}.

\begin{table}[h!]
\centering
\caption{Coefficients of the NLO contribution to the self-energy of the octet baryons (Eq.~\eqref{mass2}).}
\label{tree2coef}
~~\\[0.1em]
\begin{tabular}{cllll}
\hline\hline
    & $N$ & $\Lambda$ & $\Sigma$ & $\Xi$ \\
  \hline
  $\xi^{(a)}_{B,\pi}$ & $-(2b_0+4b_F)$ & $\frac{-2}{3}(3b_0-2b_D)$ & $-(2b_0+4b_D)$ & $-(2b_0-4b_F)$ \\
  $\xi^{(a)}_{B, K}$ & $-(4b_0+4b_D-4b_F)$ & $\frac{-2}{3}(6b_0+8b_D)$ & $-4b_0$ & $-(4b_0+4b_D+4b_F)$ \\
\hline\hline
\end{tabular}
\end{table}

At $\mathcal{O}(p^3)$  diagram Fig.~\ref{feydia}(b) gives the NLO SU(3)-breaking corrections to the baryon masses
\begin{equation}\label{mass3}
  m_{B}^{(3)}=\frac{1}{(4\pi F_\phi)^2}\sum\limits_{\phi=\pi,~K,~\eta}\xi_{B,\phi}^{(b)}H_B^{(b)}(M_{\phi}).
\end{equation}
The coefficients $\xi_{B,\phi}^{(b)}$ are given in Table~\ref{nlocoef}, and the corresponding loop functions can be found in Ref.~\cite{MartinCamalich2010_Phys.Rev.D82_074504}.

\begin{table}[h!]
\centering
\caption{Coefficients of the NNLO contribution to the self-energy of the octet baryons (Eq.~\eqref{mass3}).}
\label{nlocoef}
~~\\[0.1em]
\begin{tabular}{cllll}
\hline\hline
    & $N$ & $\Lambda$ & $\Sigma$ & $\Xi$ \\
  \hline
  $\xi^{(b)}_{B,\pi}$ & $\frac{3}{2}(D+F)^2$ & $2D^2$ & $\frac{2}{3}(D^2+6F^2)$ & $\frac{3}{2}(D-F)^2$ \\
  $\xi^{(b)}_{B,K}$   & $\frac{1}{3}(5D^2-6DF+9F^2)$  & $\frac{2}{3}(D^2+9F^2)$ & $2(D^2+F^2)$ & $\frac{1}{3}(5D^2+6DF+9F^2)$\\
  $\xi^{(b)}_{B,\eta}$ & $\frac{1}{6}(D-3F)^2$ & $\frac{2}{3}D^2$ & $\frac{2}{3}D^2$ & $\frac{1}{6}(D+3F)^2$\\
\hline\hline
\end{tabular}
\end{table}

The NNLO SU(3)-breaking corrections to the octet baryon masses are
\begin{eqnarray}\label{mass4}
  m_B^{(4)}&=&\xi_{B,\pi}^{(c)}M_{\pi}^4~+\xi_{B,K}^{(c)}M_{K}^4~+
  \xi_{B,\pi K}^{(c)}M_{\pi}^2M_{K}^2\nonumber\\
  \qquad && +\frac{1}{(4\pi F_{\phi})^2}\sum\limits_{\phi=\pi,~K,~\eta} \left[\xi_{B,\phi}^{(d,1)}H_B^{(d,1)}(M_{\phi}) +\xi_{B,\phi}^{(d,2)}H_B^{(d,2)}(M_{\phi})+\xi_{B,\phi}^{(d,3)}H_B^{(d,3)}(M_{\phi})\right]\nonumber\\
  \qquad && +\frac{1}{(4\pi F_{\phi})^2}\sum_{\substack{\phi=\pi,~K,~\eta\\  B^{\prime}=N,~\Lambda,~\Sigma,~\Xi}}\xi^{(e)}_{BB^{\prime},\phi}\cdot H_{B,B^{\prime}}^{(e)}(M_{\phi}).
\end{eqnarray}
The first three terms of Eq. \eqref{mass4}  are the tree contributions  of diagram Fig.~\ref{feydia}(c), and the corresponding coefficients $\xi_{B,\pi}^{(c)}$, $\xi_{B,K}^{(c)}$, $\xi_{B,\pi K}^{(c)}$ can be found in Table~\ref{treecoef}. The next term is the contribution from the tadpole diagram Fig.~\ref{feydia}(d) and the Clebsch-Gordan coefficients are listed in Table~\ref{tadpolecof}. The last term is from the
one-loop diagram of Fig.~\ref{feydia}(e), together with the wave function renormalization diagrams not shown,  and $\xi^{(e)}_{BB^{\prime},\phi}$ can be found in Table~\ref{loopcoef}. The loop functions, after we use the $\overline{\rm MS}$ renormalization scheme to remove the divergent pieces and the EOMS renormalization scheme to remove the PCB terms,  are written as
\begin{eqnarray}
  H_B^{(d,1)}(M_{\phi})&=&M_{\phi}^2\left[1+{\rm ln}\left(\frac{\mu^2}{M_{\phi}^2}\right)\right],\\
  H_B^{(d,2)}(M_{\phi})&=&M_{\phi}^4\left[1+{\rm ln}\left(\frac{\mu^2}{M_{\phi}^2}\right)\right],\\
  H_B^{(d,3)}(M_{\phi})&=&m_0\left\{\frac{M_{\phi}^4}{4}\left[1+{\rm ln}\left(\frac{\mu^2}{M_{\phi}^2}\right)\right]+\frac{1}{8}M_{\phi}^4\right\},
\end{eqnarray}
\begin{eqnarray}
  H_{B,B^{\prime}}^{(e)}(M_{\phi})&=&\frac{2M_{\phi}^3}{m_0^2\sqrt{4m_0^2-M_{\phi}^2}} \left[6m_0^2(m_{B}^{(2)}-m_{B^{\prime}}^{(2)})-M_{\phi}^2(2m_{B}^{(2)}-m_{B^{\prime}}^{(2)})\right] \arccos\frac{M_{\phi}}{2m_0}\nonumber\\
  \qquad && -M_{\phi}^2\left[3(m_{B}^{(2)}-m_{B^{\prime}}^{(2)})+ \frac{3m_0^2(m_{B}^{(2)}-m_{B^{\prime}}^{(2)}) -M_{\phi}^2(2m_{B}^{(2)}-m_{B^{\prime}}^{(2)})}{m_0^2} \ln\frac{M_{\phi}^2}{m_0^2}\right.\nonumber\\
  \qquad && \qquad \quad +\left. (m_{B}^{(2)}+m_{B^{\prime}}^{(2)})\ln\frac{m_0^2}{\mu^2}\right],
\end{eqnarray}
where the $m_{B}^{(2)}$ and $m_{B^{\prime}}^{(2)}$ are the corresponding LO SU(3) corrections to the octet baryon masses given in Eq. \eqref{mass2}.

\begin{table}[h!]
\scriptsize
\centering
\caption{Coefficients of the N$^3$LO tree contribution to the self-energy of the octet baryons (Eq.~\eqref{mass4}).}
\label{treecoef}
~~\\[0.1em]
\begin{tabular}{cllll}
\hline\hline
    & $N$ & $\Lambda$ & $\Sigma$ & $\Xi$ \\
  \hline
  $\xi^{(c)}_{B,\pi}$ & $-4(4d_1+2d_5+d_7+3d_8)$ & $-4(4d_3+\frac{8}{3}d_4+d_7+3d_8)$ & $-4(4d_3+d_7+3d_8)$ & $-4(4d_1-2d_5+d_7+3d_8)$ \\
  \multirow{2}{*}{$\xi^{(c)}_{B,K}$} & $-16\left(d_1-d_2+d_3\right.$ & \multirow{2}{*}{$-16(\frac{8}{3}d_3+\frac{2}{3}d_4+d_7+d_8)$} & \multirow{2}{*}{$-16(d_7+d_8)$} & $-16\left(d_1+d_2+d_3\right.$ \\
     & $\left.\qquad-d_5+d_7+d_8\right)$ &  &  & $\left.\qquad+d_5+d_7+d_8\right)$ \\
  \multirow{2}{*}{$\xi^{(c)}_{B,\pi K}$} & $~8\left(4d_1-2d_2-d_5\right.$ & \multirow{2}{*}{$~16(\frac{8}{3}d_3+\frac{4}{3}d_4-d_7+d_8)$} & \multirow{2}{*}{$-16(d_7-d_8)$} & $8\left(4d_1+2d_2+d_5\right.$ \\
     & $\left.\qquad-2d_7+2d_8\right)$ &   &  & $\left.\qquad-2d_7+2d_8\right)$\\
\hline\hline
\end{tabular}
\end{table}

\begin{table}[h!]
\small
\centering
\caption{Coefficients of the tadpole contribution to the self-energy of the octet baryons (Eq.~\eqref{mass4}).}
\label{tadpolecof}
~~\\[0.1em]
\begin{tabular}{cllll}
\hline\hline
&$N$&$\Lambda$&$\Sigma$&$\Xi$\\
\hline
$\xi_{B,\pi}^{(d,1)}$ & $-3(2b_0+b_D+b_F)m_{\pi}^2$ & $-2(3b_0+b_D)m_{\pi}^2$ & $-6(b_0+b_D)m_{\pi}^2$ & $-3(2b_0+b_D-b_F)m_{\pi}^2$ \\
$\xi_{B,K}^{(d,1)}$ & $-2(4b_0+3b_D-b_F)m_K^2$ & $-\frac{4}{3}(6b_0+5b_D)m_K^2$ & $-4(2b_0+b_D)m_K^2$ & $-2(4b_0+3b_D+b_F)m_K^2$ \\
\multirow{2}{*}{$\xi_{B,\eta}^{(d,1)}$} & $-\frac{1}{3}\left[8(b_0+b_D-b_F)m_K^2\right.$ & $-\frac{2}{9}\left[4(3b_0+4b_D)m_K^2\right.$ & $-\frac{2}{3}\left[4b_0m_K^2\right.$ & $-\frac{1}{3}\left[8(b_0+b_D+b_F)m_K^2\right.$  \\
 & $\left.-(2b_0+3b_D-5b_F)m_{\pi}^2\right]$ & $\left.-(3b_0+7b_D)m_{\pi}^2\right]$ & $\left.+(b_D-b_0)m_{\pi}^2\right]$ & $\left.-(2b_0+3b_D+5b_F)m_{\pi}^2\right]$\\
\hline
$\xi_{B,\pi}^{(d,2)}$ & $3(b_1+b_2+b_3+2b_4)$ & $2(2b_2+3b_4)$ & $2(4b_1+2b_2+3b_4)$ & $3(b_1+b_2-b_3+2b_4)$\\
$\xi_{B,K}^{(d,2)}$ & $2(3b_1+3b_2-b_3+4b_4)$ & $\frac{4}{3}(9b_1+b_2+6b_4)$ & $4(b_1+b_2+2b_4)$ & $2(3b_1+3b_2+b_3+4b_4)$\\
$\xi_{B,\eta}^{(d,2)}$ & $\frac{1}{3}(9b_1+b_2-3b_3+6b_4)$ & $2(2b_2+b_4)$ & $\frac{2}{3}(2b_2+3b_4)$ & $\frac{1}{3}(9b_1+b_2+3b_3+6b_4)$ \\
\hline
$\xi_{B,\pi}^{(d,3)}$ & $6(b_5+b_6+b_7+2b_8)$ & $4(2b_7+3b_8)$ & $4(4b_5+2b_7+3b_8)$ & $6(b_5-b_6+b_7+2b_8)$ \\
$\xi_{B,K}^{(d,3)}$ & $4(3b_5-b_6+3b_7+4b_8)$ & $\frac{8}{3}(9b_5+b_7+6b_8)$ & $8(b_5+b_7+2b_8)$ & $4(3b_5+b_6+3b_7+4b_8)$\\
$\xi_{B,\eta}^{(d,3)}$ & $\frac{2}{3}(9b_5-3b_6+b_7+6b_8)$ & $4(2b_7+b_8)$ & $\frac{4}{3}(2b_7+3b_8)$ & $\frac{2}{3}(9b_5+3b_6+b_7+6b_8)$ \\
  \hline \hline
\end{tabular}
\end{table}

\begin{table}[h!]
\small
\centering
\caption{Coefficients of the loop contributions (Fig.~\ref{feydia}e) to the self-energy of the octet baryons (Eq.~\eqref{mass4}).}
\label{loopcoef}
~~\\[0.1em]
\begin{tabular}{llll}
\hline\hline
$N$ &$\Lambda$&$\Sigma$&$\Xi$\\
\hline
$\xi_{NN\pi}^{(e)}=\frac{3}{4}(D+F)^2$ & $\xi_{\Lambda N K}^{(e)}=\frac{1}{6}(D+3F)^2$ & $\xi_{\Sigma N K}^{(e)}=\frac{1}{2}(D-F)^2$ & $\xi_{\Xi\Lambda K}^{(e)}=\frac{1}{12}(D-3F)^2$ \\
$\xi_{NN\eta}^{(e)}=\frac{1}{12}(D-3F)^2$ & $\xi_{\Lambda\Lambda\eta}^{(e)}=\frac{1}{3}D^2$ & $\xi_{\Sigma\Lambda\pi}^{(e)}=\frac{1}{3} D^2$ & $\xi_{\Xi\Sigma K}^{(e)}=\frac{3}{4}(D+F)^2$ \\
$\xi_{N\Lambda K}^{(e)}=\frac{1}{12}(D+3F)^2$ & $\xi_{\Lambda\Sigma\pi}^{(e)}=D^2$  & $\xi_{\Sigma\Sigma\pi}^{(e)}=2F^2$ & $\xi_{\Xi\Xi\pi}^{(e)}=\frac{3}{4}(D-F)^2$ \\
$\xi_{N\Sigma K}^{(e)}=\frac{3}{4}(D-F)^2$ & $\xi_{\Lambda\Xi K}^{(e)}=\frac{1}{6}(D-3F)^2$ & $\xi_{\Sigma\Sigma\eta}^{(e)}=\frac{1}{3}D^2$ & $\xi_{\Xi\Xi\eta}=\frac{1}{12}(D+3F)^2$ \\
  &  & $\xi_{\Sigma\Xi K}^{(e)}=\frac{1}{2}(D+F)^2$ &   \\
\hline\hline
\end{tabular}
\end{table}

At N$^3$LO, a replacement of the meson masses by their $\mathcal{O}(p^4)$ counterparts in $m_B^{(2)}$ generates
N$^3$LO contributions to $m_B^{(4)}$. The corresponding Nambu-Goldstone boson masses up to $\mathcal{O}(p^4)$ can be
found in Ref.~\cite{Gasser1985_Nucl.Phys.B250_465--516}, which read as
\begin{eqnarray}
  M_{\pi, 4}^2 &=& M_{\pi, 2}^2\left\{ 1 + \frac{M_{\pi,
        2}^2}{32\pi^2F_{\phi}^2}\ln\left(\frac{M_{\pi, 2}^2}{\mu^2}\right)
    - \frac{M_{\eta, 2}^2}{96\pi^2F_{\phi}^2}\ln\left(\frac{M_{\eta,
          2}^2}{\mu^2}\right)\right.\nonumber\\
\qquad && \left.+\frac{16}{F_{\phi}^2}\left[\left(\frac{M_{\pi,2}^2}{2}+M_{K,2}^2\right)(2L^r_6-L^r_4) +\frac{M_{\pi,2}^2}{2}(2L^r_8-L^r_5)\right]\right\},
\end{eqnarray}

\begin{eqnarray}
  M_{K, 4}^2 &=& M_{K, 2}^2\left\{1+\frac{M^2_{\eta,
        2}}{48\pi^2F_{\phi}^2}\ln\left(\frac{M_{\eta,
          2}^2}{\mu^2}\right)\right.\nonumber\\
\qquad && \left.+\frac{16}{F_{\phi}^2}\left[\left(\frac{M_{\pi,2}^2}{2}+M_{K,2}^2\right)(2L^r_6-L^r_4) + \frac{M_{K,2}^2}{2}(2L^r_8-L^r_5)\right]\right\},
\end{eqnarray}

\begin{eqnarray}
  M_{\eta, 4}^2 &=& M_{\pi, 2}^2\left[\frac{M_{\eta,
        2}^2}{96\pi^2F_{\phi}^2}\ln\left(\frac{M_{\eta,
          2}^2}{\mu^2}\right)-\frac{M_{\pi,
        2}^2}{32\pi^2F_{\phi}^2}\ln\left(\frac{M_{\pi, 2}^2}{\mu^2}\right)
    + \frac{M_{K, 2}^2}{48\pi^2F_{\phi}^2}\ln\left(\frac{M_{K,
          2}^2}{\mu^2}\right)\right]\nonumber\\
\qquad && + M_{\eta, 2}^2\left[1+\frac{M_{K,
      2}^2}{16\pi^2F_{\phi}^2}\ln\left(\frac{M_{K,
        2}^2}{\mu^2}\right)-\frac{M_{\eta,
      2}}{24\pi^2F_{\phi}^2}\ln\left(\frac{M_{\eta,
        2}}{\mu^2}\right)\right.\nonumber\\
\qquad && \left. +\frac{16}{F_{\phi}^2}\left(\frac{M_{\pi,2}^2}{2}+M_{K,2}^2\right)(2L^r_6-L^r_4)+
  8\frac{M_{\eta, 2}^2}{F_{\phi}^2}(2L^r_8-L^r_5)\right]\nonumber\\
\qquad && +\frac{128}{9}\frac{(M_{K,2}^2-M_{\pi,2}^2)^2}{F_{\phi}^2}(3L^r_7+L^r_8).
\end{eqnarray}
The empirical values of $2L^r_6-L^r_4=-0.17\times10^{-3}$ and $2L^r_8-L^r_5=-0.22\times 10^{-3}$, and $3L^r_7+L^r_8=-0.15\times 10^{-3}$ are taken from the latest global fit~\cite{Bijnens:2011tb}, which are evaluated at the renormalization scale $\mu=0.77$ GeV.~\footnote{ It should be noted that the uncertainties of the $L^r_i$ are quite large. Because the effects of their contributions are found to be small, we do not take into account the uncertainties of these LECs in our fit of the LQCD mass data.} To be consistent with our renormalization scale used
for the one-baryon sector, we have re-evaluated the $L^r_i$'s at $\mu=1$ GeV.

\subsection{Finite-volume corrections}
Because lattice QCD simulations are performed in a finite hypercube, the momenta of virtual particles
are discretized. As a result, the simulated results are different from those of the infinite space-time.
The difference is termed as finite-volume corrections (FVCs). In cases where $M_{\phi} L\gg1$, the so-called $p$-regime, ChPT
provides a model-independent framework to study FVCs. In the past, most studies have employed either heavy-baryon ChPT or infrared BChPT and focused on the two flavor sector (see, e.g., Refs.~\cite{Khanetal.(QCDSF-UKQCDCollaboration)2003_Nucl.Phys.B689_175,Procura2006_Phys.Rev.D73_114510}). Recently, two studies in the three-flavor sector have been performed with HB ChPT~\cite{Beaneitetal.(NPLQCDCollaboration)2011_PhysicalReviewD84_014507} and EOMS BChPT~\cite{Geng2011_Phys.Rev.D84_074024}. It was
pointed out that FVCs computed with infrared BChPT and EOMS BChPT are the same. Furthermore, it is noted that
both HB ChPT and EOMS BChPT can describe the NPLQCD data, with different values for the $N\Delta\pi$  coupling. However, it was also
noted that with the LECs determined from the volume dependence of the LQCD data, EOMS BChPT extrapolation to the physical point are in better agreement with data.

The $\mathcal{O}(p^3)$ covariant BChPT results can be found in Ref.~\cite{Geng2011_Phys.Rev.D84_074024}.
 Following the same procedure, one can easily calculate the $\mathcal{O}(p^4)$ results in a finite hypercube
  by replacing the $H$'s of Eq.~(\ref{mass4}) by $\tilde{H}=H+\delta G$ with

\begin{equation}
  \delta G_B^{(d,1)}(M_\phi)=\frac{1}{2}\delta_{1/2}\left(M_\phi^2\right),
\end{equation}
\begin{equation}
  \delta G_B^{(d,2)}(M_\phi)=\frac{M_\phi^2}{2}\delta_{1/2}\left(M_\phi^2\right),
\end{equation}
\begin{equation}
  \delta G_B^{(d,3)}(M_\phi)=\frac{m_0}{2}\delta_{-1/2}\left(M_\phi^2\right),
\end{equation}
\begin{eqnarray}
  \delta G_{B,B^{\prime}}^{(e)}&=&\int_0^1dx\left\{-\frac{1}{2}(2xm_{B}^{(2)}+m_{B^{\prime}}^{(2)})
  \delta_{1/2}(\mathcal{M}_{N}^2)\right.\nonumber\\
  \qquad && -\frac{1}{4}\left[M_{\phi}^2(x-1)\left(m_{B}^{(2)}(1+x)+m_{B^{\prime}}^{(2)}\right) -2xm_0^2\left(m_{B}^{(2)}(5x^2-3)+m_{B^{\prime}}^{(2)}(4x+3)\right)\right]
  \delta_{3/2}(\mathcal{M}_{N}^2)\nonumber\\
  \qquad &&\left. -\frac{1}{4}\left[6m_0^4(x+1)x^3\left(m_{B}^{(2)}(x-1)+m_{B^{\prime}}^{(2)}\right)\right.\right.\nonumber\\
  \qquad && \left.\left.\qquad -3m_0^2M_{\phi}^2x(x-1)(x+2)\left(m_{B}^{(2)}(x-1)+m_{B^{\prime}}^{(2)}\right)\right] \delta_{5/2}(\mathcal{M}_{N}^2)\right\},
\end{eqnarray}
where $\mathcal{M}_N^2=x^2m_0^2+(1-x)M_{\phi}^2-i\epsilon$, and
\begin{equation}
\delta_r (\mathcal{M}^2)=
\frac{2^{-1/2-r}(\sqrt{\mathcal{M}^2})^{3-2r}}{\pi^{3/2}\Gamma(r)}\sum_{\vec{n}\ne0}(L\sqrt{\mathcal{M}^2}|\vec{n}|)^{-3/2+r}K_{3/2-r}(L\sqrt{\mathcal{M}^2}|\vec{n}|)
\end{equation}
 with $K_n(z)$ the modified Bessel function of the second kind, and $\sum\limits_{\vec{n}\ne0}\equiv\sum\limits^{\infty}_{n_x=-\infty}\sum\limits^{\infty}_{n_y=-\infty}\sum\limits^{\infty}_{n_z=-\infty}(1-\delta(|\vec{n}|,0))$ with $\vec{n}=(n_x,n_y,n_z)$.

\section{Chiral extrapolation of octet baryon masses}\label{sec4}
In this section, we study light quark mass dependence of the g.s. octet baryon masses using the N$^3$LO EOMS-BChPT mass formulas by fitting the $n_f=2+1$ LQCD simulation results. As mentioned above, there are 19 unknown LECs, which cannot be fully determined by the lattice data of a single LQCD simulation. Therefore, we decide to fit all the publicly available lattice results of the g.s. octet baryon masses obtained by different collaborations, including the PACS-CS~\cite{Aokiitetal.(PACS-CSCollaboration)2009_Phys.Rev.D79_034503}, the LHPC~\cite{Walker-Louditelal.2009_Phys.Rev.D79_054502}, the HSC~\cite{Linitetal.(HSCCollaboration)2009_Phys.Rev.D79_034502}, the QCDSF-UKQCD~\cite{Bietenholzitetal.(QCDSF-UKQCDCollaboration)2011_PhysicalReviewD84_054509} and the NLPQCD~\cite{Beaneitetal.(NPLQCDCollaboration)2011_PhysicalReviewD84_014507} collaborations. In performing such a study, we can test the consistency of all these lattice simulations, which used totally different setup. Furthermore, because different calculations are performed with different lattice spacing, the uncertainties of our results also incorporate (partly) discretization effects.

\subsection{2+1 flavor LQCD data of octet baryon masses}

\begin{figure}[t]
\centering
\includegraphics[width=15cm]{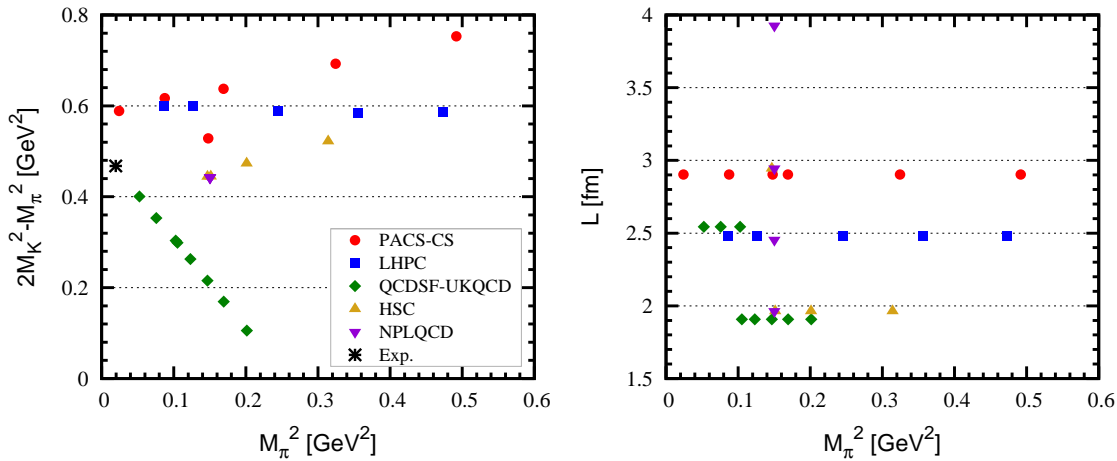}
\caption{(Color online). Landscape of LQCD simulations of the ground-state octet baryon masses: the PACS-CS (red circles),  LHPC (blue squares), QCDSF-UKQCD (green diamonds), HSC (yellow upper triangles) and NPLQCD (purple lower triangles) Collaborations in the $2M_K^2-M_{\pi}^2$ vs. $M_{\pi}^2$ plane (left panel) and in the $L$ vs. $M_{\pi}^2$ plane (right panel). The star denotes the physical point with the physical light- and strange-quark masses \textbf{(as implied by leading-order ChPT)}.}
\label{fig2}
\end{figure}

As has been stated in the introduction, the lattice calculations are performed with larger than physical light quark masses and finite volumes. In Fig.~\ref{fig2} , we  show the lattice simulation points of the PACS-CS, LHPC, HSC, QCDSF-UKQCD and NPLQCD collaborations in the ($2M_K^2-M_{\pi}^2$) --$M_{\pi}^2$ plane and in the $L$--$M_{\pi}^2$ plane. There are only one point for the NPLQCD simulation at $M_{\pi}=389$ MeV on the left panel, because the main purpose of the NPLQCD collaboration is to study the finite-volume effects on the baryon masses. The large range of light pion masses provides an opportunity to explore the applicability of ChPT for extrapolation of baryon masses. Although the light $u/d$ quark masses adopted are always larger than their physical counterpart, the strange quark masses vary from collaboration to collaboration: those of the PACS-CS and LHPC collaborations are larger than the physical one; those of the HSC and NPLQCD groups are a bit smaller, whereas those of the QCDSF-UKQCD collaboration are all lighter than the physical one.

In the $L$--$M_{\pi}^2$ plane, it is seen that the PACS-CS and LHPC simulations adopt a single value of lattice volume; the HSC and QCDSF-UKQCD simulations use two different lattice volumes and the NPLQCD simulations are performed with four different lattice volumes in order to study the finite-volume effects on the octet baryon masses. Many of the simulations are still performed with $M_{\phi}L$ from $3$ to $5$ and with $M_{\phi}$ larger than $300$ MeV. As a result, FVCs may not be negligible (see, e.g., Ref.~\cite{Geng2011_Phys.Rev.D84_074024}). In our study, we will take into account FVCs in a self-consist way as in Refs.~\cite{MartinCamalich2010_Phys.Rev.D82_074504,Geng2011_Phys.Rev.D84_074024}.  It should be stressed that taking into account FVCs is not only needed to obtain a good fit of the LQCD data but also can constrain better the values of some of the LECs, since they contribute differently to FVCs and to light quark mass dependence.

Except for varying light- and strange-quark masses and lattice size,  lattice simulations can adopt different fermion/gauge actions, all of which are believed to lead to the same continuum theory.  Therefore, it's crucial to test whether all these simulation results are consistent with each other \cite{Beringeritetal.(ParticleDataGroup)2012_Phys.Rev.D86_010001}.

In Appendix \ref{latticedata}, we tabulate the octet  baryon masses of the PACS-CS, LHPC, HSC, QCDSF-UKQCD and NPLQCD collaborations. The numbers are given in physical units using either the lattice scale specified in the original publications~\cite{Aokiitetal.(PACS-CSCollaboration)2009_Phys.Rev.D79_034503, Walker-Louditelal.2009_Phys.Rev.D79_054502, Linitetal.(HSCCollaboration)2009_Phys.Rev.D79_034502, Beaneitetal.(NPLQCDCollaboration)2011_PhysicalReviewD84_014507} or the method of ratios such as QCDSF-UKQCD~\cite{Bietenholzitetal.(QCDSF-UKQCDCollaboration)2011_PhysicalReviewD84_054509}.  N$^3$LO BChPT is not supposed to be able to describe all the lattice simulations, which can have very large light/strange quark masses and quite small volumes. Recent studies show that $M_\pi< 0.4-0.5$ GeV may be already at the limit where N$^3$LO BChPT can work (see, e.g., Ref.~\cite{Ohkiitetal.(JLQCDCollaboration)2009_ProceedingsofScienceLAT2009_124}). Therefore,  to reduce contributions from higher order terms in the chiral expansion, we choose the lattice simulations with  $M_{\pi}^2<0.25$ GeV$^2$.  To minimize FVCs, we further require the lattice data to have $M_{\phi}L>4$ .  There are only 11 lattice data sets satisfying both requirements: three  from PACS-CS, two from LHPC, only one from QCDSF-UKQCD, two from HSC and three from NPLQCD. We denote these lattice sets with stars in Tables~\ref{PACS-CSdata} -- \ref{NPLQCDdata} of Appendix \ref{latticedata}. For comparison, we also choose a large data set with $M_{\pi}^2<0.5$ GeV$^2$, $M_{K}^2<0.7$ GeV$^2$ and $M_{\phi}L>3$.  There are 26 lattice data sets left and they are listed in Table~\ref{PACS-CSdata} (six sets  from PACS-CS), Table~\ref{LHPCdata} (five sets from LHPC), Table~\ref{HSCdata} (four sets from HSC), Table~\ref{UKQCDdata} (eight sets from QCDSF-UKQCD) and Table~\ref{NPLQCDdata} (four sets from NPLQCD). In the following, we denote the above 11 lattice sets as Set-I, and the 26 sets as Set-II. It should be mentioned that we are well aware that N$^3$LO BChPT may not be able to describe the larger data set, which is indeed the case as we will see in the following subsection.

\subsection{Chiral extrapolation of octet baryon masses}
We proceed to fit the LQCD results of octet baryon masses by using the N$^3$LO EOMS-BChPT mass formulas Eq.~\eqref{chiexpand}. For the meson decay constant, we use $F_\phi=0.0871$ GeV~\cite{Amoros2001_Nucl.Phys.B602_87}. In principle at N$^3$LO one can use either the
chiral limit value $F_\phi=0.0871$ GeV obtained from a two-loop ChPT calculation~\cite{Amoros2001_Nucl.Phys.B602_87}, or the SU(3) averaged value, $F_\phi=1.17 F_\pi$ with $F_\pi=0.0924$ GeV  as in Ref.~\cite{MartinCamalich2010_Phys.Rev.D82_074504}. The difference is of higher chiral order. In practice, we found that at N$^3$LO the results
are not sensitive to these two options while at NNLO the SU(3) averaged value is more preferred by the LQCD data. For the baryon axial coupling constants $D$ and $F$, we use the SU(6) relation $F=2/3 D$. Together with $D+F=1.26$ as determined from nuclear beta decay,  one then has $D=0.8$ and $F=0.46$. We have allowed  $D$ and $F$ to vary in the fits and found that the optimal values determined by the lattice data are consistent with the phenomenological values. The renormalization scale $\mu$ is set at 1 GeV, as in Ref.~\cite{MartinCamalich2010_Phys.Rev.D82_074504}.

Set-I and Set-II contain 11 and 26 sets of data from the PACS-CS, LHPC, HSC, QCDSF-UKQCD and NPLQCD collaborations, respectively. The results of different collaborations are not correlated with each other, but the data form the same collaboration are partially correlated by the uncertainties propagated from the determination of the lattice spacing. Therefore, in order to correctly calculate the $\chi^2$ we incorporate the inverse of the resulting correlation matrix $C_{ij}=\sigma_i\sigma_j\delta_{ij}+\Delta a_i\Delta a_j$ for each lattice ensemble (see Ref.~\cite{MartinCamalich2010_Phys.Rev.D82_074504})
, where the $\sigma_i$ are the lattice statistical errors and the $\Delta a_i$ are the fully-correlated errors propagated from the determination of $a_i$.  The FVCs to the baryon masses are consistently calculated in the EOMS-BChPT framework as explained
in section 3.4.

\begin{table}[t]
\centering
\caption{Values of the LECs from the best fit to the LQCD data and the experimental data at $\mathcal{O}(p^2)$, $\mathcal{O}(p^3)$, and $\mathcal{O}(p^4)$. The estimator for the fits $\chi^2$/d.o.f. is also given (see text for details).}
\label{octetfitcoef}
~~\\[0.1em]
\begin{tabular}{lrrr|r}
\hline\hline
         & \multicolumn{3}{c}{Set-I} & Set-II \\
      \cline{2-5}
         & Fit - $\mathcal{O}(p^2)$ & Fit - $\mathcal{O}(p^3)$  &   Fit I - $\mathcal{O}(p^4)$   &  Fit II - $\mathcal{O}(p^4)$  \\
\hline
  $m_0$~[MeV]         & $900(6)$      &   $767(6)$     &  $880(22)$      & $868(12)$     \\
  $b_0$~[GeV$^{-1}$]  &$-0.273(6)$    &  $-0.886(5)$   &  $-0.609(19)$   & $-0.714(21)$  \\
  $b_D$~[GeV$^{-1}$]  &$0.0506(17)$   &  $0.0482(17)$  &  $0.225(34)$    & $0.222(20)$   \\
  $b_F$~[GeV$^{-1}$]  &$-0.179(1)$    &  $-0.514(1)$   &  $-0.404(27)$   & $-0.428(12)$  \\
  $b_1$~[GeV$^{-1}$]  & --            &  --            &  $0.550(44)$    & $0.515(132)$   \\
    $b_2$~[GeV$^{-1}$]  & --            &  --            &  $-0.706(99)$   & $0.148(48)$   \\
  $b_3$~[GeV$^{-1}$]  & --            &  --            &  $-0.674(115)$  & $-0.663(155)$   \\
  $b_4$~[GeV$^{-1}$]  & --            &  --            &  $-0.843(81)$   & $-0.868(105)$    \\
  $b_5$~[GeV$^{-2}$]  & --            &  --            &  $-0.555(144)$  & $-0.643(246)$ \\
  $b_6$~[GeV$^{-2}$]  & --            &  --            &  $0.160(95)$    & $-0.268(334)$ \\
  $b_7$~[GeV$^{-2}$]  & --            &  --            &  $1.98(18)$     & $0.176(72)$  \\
  $b_8$~[GeV$^{-2}$]  & --            &  --            &  $0.473(65)$    & $-0.0694(1638)$ \\
  $d_1$~[GeV$^{-3}$]  & --            &  --            &  $0.0340(143)$  & $0.0345(134)$  \\
  $d_2$~[GeV$^{-3}$]  & --            &  --            &  $0.296(53)$    & $0.374(21)$    \\
  $d_3$~[GeV$^{-3}$]  & --            &  --            &  $0.0431(304)$  & $0.00499(1817)$ \\
  $d_4$~[GeV$^{-3}$]  & --            &  --            &  $0.234(67)$    & $0.267(34)$     \\
  $d_5$~[GeV$^{-3}$]  & --            &  --            &  $-0.328(60)$   & $-0.445(26)$    \\
  $d_7$~[GeV$^{-3}$]  & --            &  --            &  $-0.0358(269)$ & $-0.183(12)$    \\
  $d_8$~[GeV$^{-3}$]  & --            &  --            &  $-0.107(32)$   & $-0.307(21)$    \\
  \hline
$\chi^2$/d.o.f. & $11.8$ & $8.6$  & $1.0$ & $1.6$\\
\hline\hline
\end{tabular}
\end{table}

We perform a $\chi^2$ fit to the lattice data  and the physical octet baryon masses by varying the $19$ LECs.
The so-obtained values of the LECs from the best fits are listed in Table~\ref{octetfitcoef}. We remark that in the following the experimental octet baryon masses are always included in the fit unless otherwise stated. This way, the prediction of the sigma terms of the octet baryons and the fitted values of LECs are better constrained.

For the sake of comparison, we have fitted Set-I using the NLO~\footnote{ Since at this order ChPT does not generate any FVCs, we have subtracted from the LQCD data the FVCs calculated by N$^3$LO EOMS-BChPT with the LECs determined from the corresponding best fit.} and NNLO EOMS-BChPT. The values of the LECs $b_0$, $b_D$, $b_F$, and $m_0$ are tabulated in Table~\ref{octetfitcoef}. An order-by-order improvement is clearly seen, with decreasing $\chi^2/\mathrm{d.o.f.}$ at each increasing order. Apparently, only using the $\mathcal{O}(p^3)$ chiral expansion, we cannot describe simultaneously  the lattice data from the five collaborations. The corresponding $\chi^2/{\rm d.o.f.}$ is about $8.6$. On the other hand, in the N$^3$LO fit of lattice data Set-I and experimental octet baryon masses, $\chi^2/{\rm d.o.f.} = 1.0$. In addition, the values of the fitted LECs (named Fit I) all look very natural.~\footnote{We have
checked that removing from lattice data Set-I the two lattice points of LHPC and HSC with $M_\pi>400$ MeV and $M_K>580$ MeV  does not change qualitatively
our results.} Especially, the baryon mass in the chiral limit $m_0 = 880$ MeV seems to be consistent with the SU(2)-BChPT  value \cite{Procura2003_Phys.Rev.D69_034505,Procura2006_Phys.Rev.D73_114510}. The fit to data Set-II yields a $\chi^2/{\rm d.o.f.}$ about $1.6$ and the fitted LECs look similar to those from Fit I except
$b_2$, $b_6$, $b_7$ and $b_8$.~\footnote{ It should be noted that some of the LECs are correlated and other data except the masses are needed to
fully determine the LECs.} The increased $\chi^2$/d.o.f. indicates that data set II is a bit beyond the application region of N$^3$LO BChPT.
Finally, it is important to point out that including the FVCs is important to understand the LQCD results in ChPT at N$^3$LO. Without FVCs taken into account, the best fit to lattice data Set-I yields  $\chi^2/\mathrm{d.o.f.} \sim 1.9$.
\begin{figure}[t]
\centering
  \includegraphics[width=15cm]{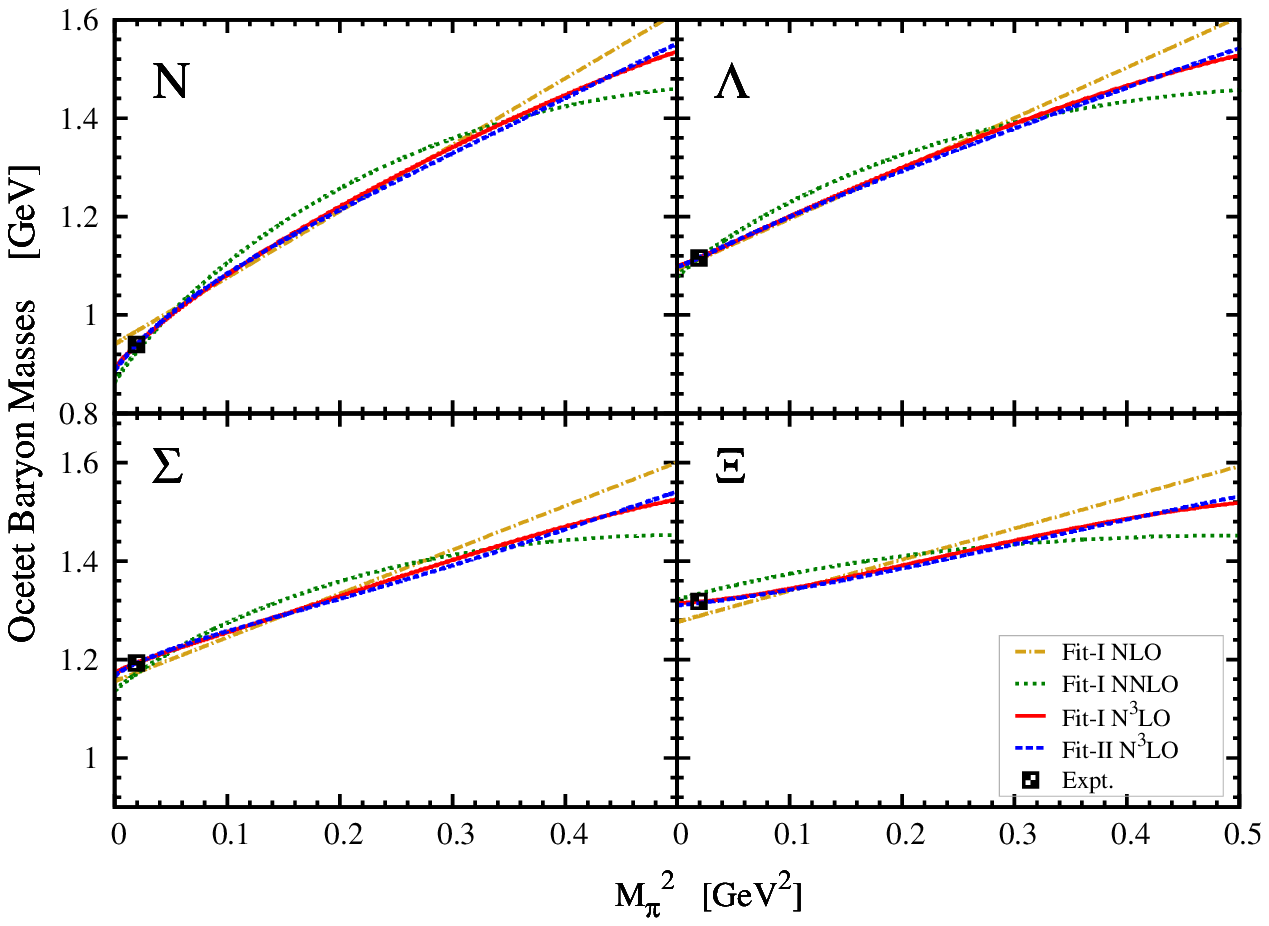}\\
  \caption{(Color online). The lowest-lying baryon octet masses as functions of the pion mass.
   The dot-dashed lines and the dotted lines are the best NLO and NNLO fits to lattice data Set-I. The red solid line and the blue dashed line correspond to the best N$^3$LO fits to lattice data Set-I and Set-II, respectively.
    In obtaining the ChPT results, the strange quark mass has been set to its physical value. 
  }
  \label{NLO-NNLO-phy}
\end{figure}

In Fig.~\ref{NLO-NNLO-phy}, setting the strange-quark mass to its physical value, we plot the light-quark mass evolution of $N$, $\Lambda$, $\Sigma$ and $\Xi$ as functions of $M_{\pi}^2$ using the LECs from Table~\ref{octetfitcoef}. We can see that the NNLO fitting results are more curved and
do not describe well lattice data Set-I. On the contrary the two N$^3$LO fits, named Fit I and Fit II, both can give a good description of lattice data Set-I. The rather linear dependence of the lattice data on $M_\pi^2$ at large light quark masses, which are exhibited both by the lattice data~\cite{Walker-Louditelal.2009_Phys.Rev.D79_054502} and
reported by other groups, is clearly seen.

\begin{figure}[t]
\centering
  \includegraphics[width=15cm]{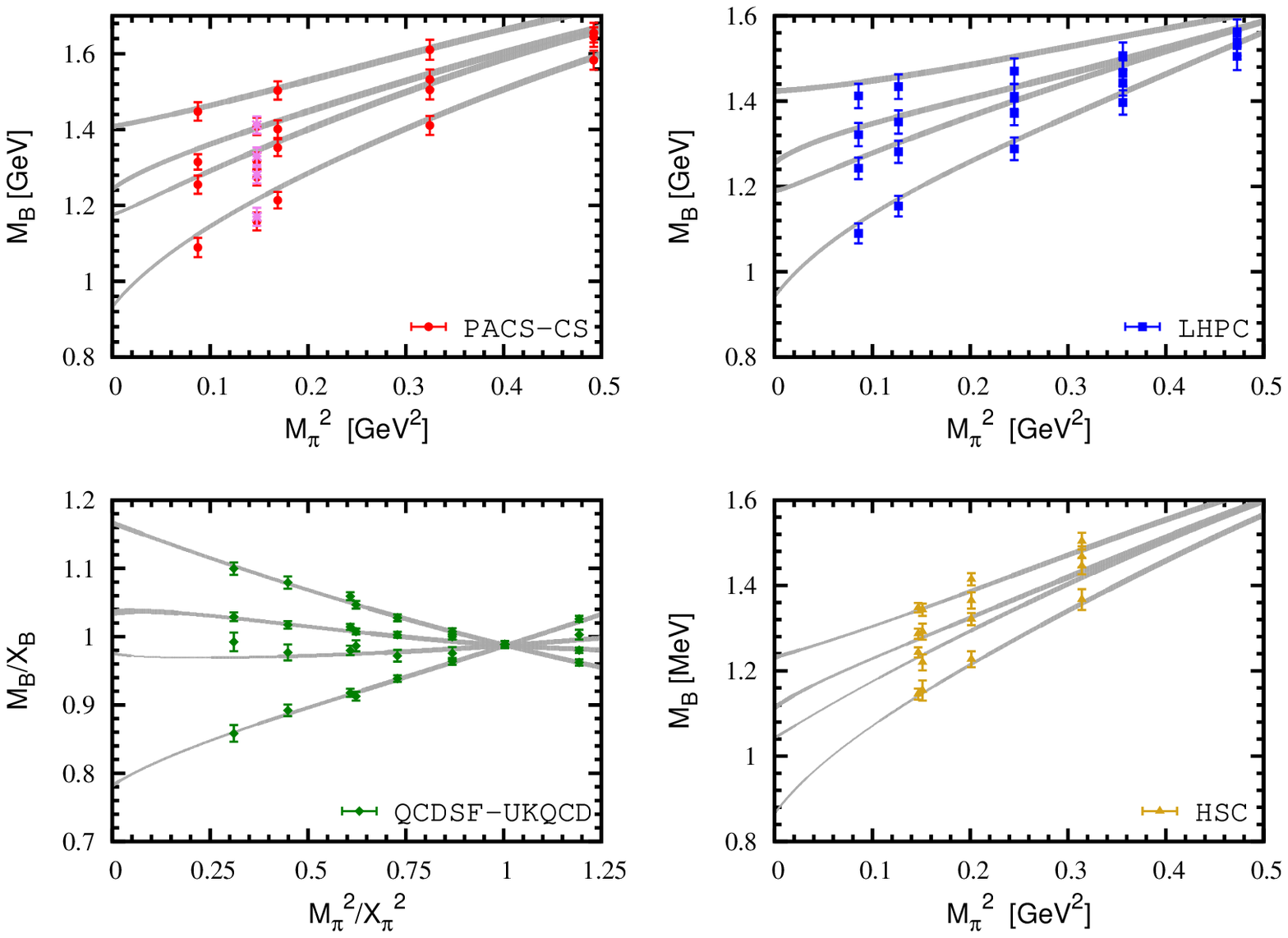}\\
  \caption{(Color online). The PACS-CS, LHPC, QCDSF-UKQCD and HSC data in comparison to the BChPT fits as functions of the pion mass.
  The curves from bottom to top are for $N$, $\Lambda$, $\Sigma$ and $\Xi$, respectively. The bands in each panel are the Fit-II results at the 68\% confidence level with $M_K^2=a+bM_\pi^2$, where $a$ and $b$ are
  determined by fitting $M_K^2$ as a linear function of $M_{\pi}^2$ for the corresponding lattice ensemble (see Appendix~\ref{app2}). It should be noted that because one of the PACS-CS data (denoted by the solid pink point) has a small strange-quark mass, it is not included in the $M_K^2$ fit. FVCs have been subtracted from the lattice data using the corresponding BChPT fit. The QCDSF-UKQCD collaboration defined $X_{\pi}=\sqrt{(M_{\pi}^2+2M_K^2)/3}$, $X_{B}=(m_N+m_{\Sigma}+m_{\Xi})/3$, where the meson and baryon masses are the physical ones.}
  \label{octetfitting}
\end{figure}

As mentioned above, Fit II is a global fit to the $26$ lattice data sets with a $\chi^2/{\rm d.o.f.}~=~1.6$. In Fig.~\ref{octetfitting}, we show its description of the lattice data from the PACS-CS, LHPC, QCDSF-UKQCD and HSC collaborations, respectively.~\footnote{ We do not show the NPLQCD data because they are obtained at a single pion mass.} The baryon masses are plotted as functions of $M_\pi^2$, with the kaon mass calculated using $M_{K}^2=a+bM_{\pi}^2$  with $a$ and $b$ determined from the lattice data for each ensemble (see Appendix~\ref{app2} for details), and the lattice data are all extrapolated to infinite space-time. It is clear that our fitting results can give a reasonable description of the lattice data, with  $\chi^2/{\rm d.o.f.}$ equals $0.93$, $1.38$, $0.93$, $0.92$ for each collaboration's data, respectively.  It should be noted that Fig.~\ref{octetfitting} is only for demonstration purposes because we should not trust the fit-II results, which might be misleading because of the inclusion of lattice data beyond the application region of N$^3$LO BChPT.

\section{Pion- and strangeness-baryon sigma terms}\label{sec5}
In this section, we evaluate the pion- and strangeness-sigma terms for the octet baryons at the physical point using the N$^3$LO BChPT results of the previous section. Baryon sigma terms are quantities of great importance. In particular, the nucleon sigma terms  are vital quantitities in understanding the composition of the nucleon mass and the strangeness content of the nucleon. An accurate knowledge of the strangeness-nucleon sigma term is also useful for dark matter searches~\cite{Giedt2009_Phys.Rev.Lett.103_201802}. ChPT, with its LECs fixed by the LQCD mass data, can make predictions for sigma terms~\cite{Durritetal.(BMWCollaboartion)2012_Phys.Rev.D85_014509,Shanahan2012__,Semke2012_Phys.Lett.B717_242}.

The sigma terms are the scalar form factors of baryons at zero recoil. In this work, we calculate all the octet baryon sigma terms $\sigma_{\pi B}$, $\sigma_{sB}$ for $B=N,~\Lambda,~\Sigma,~\Xi$ , by use of the Feynman-Hellmann theorem, which states:
\begin{eqnarray}\label{sigmacal}
  \sigma_{\pi B} &=& m_l\langle B(p)|\bar{u}u+\bar{d}d|B(p)\rangle = m_{l}\frac{\partial M_B}{\partial m_l},\\
  \sigma_{s B} &=& m_s\langle B(p)|\bar{s}s|B(p)\rangle = m_s\frac{\partial M_B}{\partial m_s},
\end{eqnarray}
where $m_l=(m_u+m_d)/2$. Using leading-order ChPT, the quark masses can be expressed by the pseudoscalar masses, with $m_l=M_{\pi}^2/(2B_0)$ and $m_s=(2M_K^2-M_{\pi}^2)/(2B_0)$.
Other related quantities, which often appear in the literature, including the strangeness content ($y_B$) and the so-called ``dimensionless sigma terms" ($f_{lB}$, $f_{sB}$), are also calculated:
\begin{eqnarray}\label{strangenesscal}
  y_B &=& \frac{2\langle B(p)|\bar{s}s|B(p)\rangle}{\langle B(p)|\bar{u}u+\bar{d}d|B(p)\rangle}=\frac{m_l}{m_s}\frac{2\sigma_{sB}}{\sigma_{\pi B}},\\
  f_{lB} &=& \frac{m_{l}\langle B(p)|\bar{u}u+\bar{d}d|B(p)\rangle}{M_B}=\frac{\sigma_{\pi B}}{M_B},\\
  f_{sB} &=& \frac{m_s\langle B(p)|\bar{s}s|B(p)\rangle}{M_B}=\frac{\sigma_{s B}}{M_B}.
\end{eqnarray}

Using the Fit-I LECs, we obtain the pion- and strangeness sigma terms $\sigma_{\pi B}$, $\sigma_{s B}$ for all the  octet baryons, and the corresponding strangeness content $y_{B}$, ``dimensionless sigma terms" $f_{lB}$, $f_{sB}$. The results are shown in Table \ref{sigmaresults}.

\begin{table}[b]
\centering
\caption{The sigma-terms, the strangeness content and the ``dimensionless sigma terms" of the octet baryons at the physical point.
The first error is statistical and the second one is systematic, estimated by taking half the difference between the N$^3$LO result and the NNLO result.}
~~\\[0.1em]
\begin{tabular}{cccccccc}
\hline\hline
       & $\sigma_{\pi B}$~[MeV]  & $\sigma_{sB}$~[MeV]  & $y_{B}$  & $f_{lB}$  & $f_{sB}$ \\
\hline
$N$       &  $43(1)(6)$  & $126(24)(54)$    &  $0.244(47)(110)$    & $0.0457(11)(64)$   & $0.134(26)(57)$ \\
$\Lambda$ &  $19(1)(7)$  & $269(23)(66)$    &  $1.179(118)(522)$   & $0.0170(9)(63)$   & $0.241(21)(59)$ \\
$\Sigma$  &  $18(2)(6)$  & $296(21)(50)$    &  $1.369(180)(512)$   & $0.0151(17)(50)$   & $0.248(18)(42)$ \\
$\Xi$     &  $ 4(2)(3)$  & $397(22)(56)$    &  $8.263(4157)(6306)$ & $0.00303(152)(228)$ & $0.301(17)(42)$ \\
\hline\hline
\end{tabular}
\label{sigmaresults}
\end{table}

\begin{figure}[t]
\centering
  \includegraphics[width=15cm]{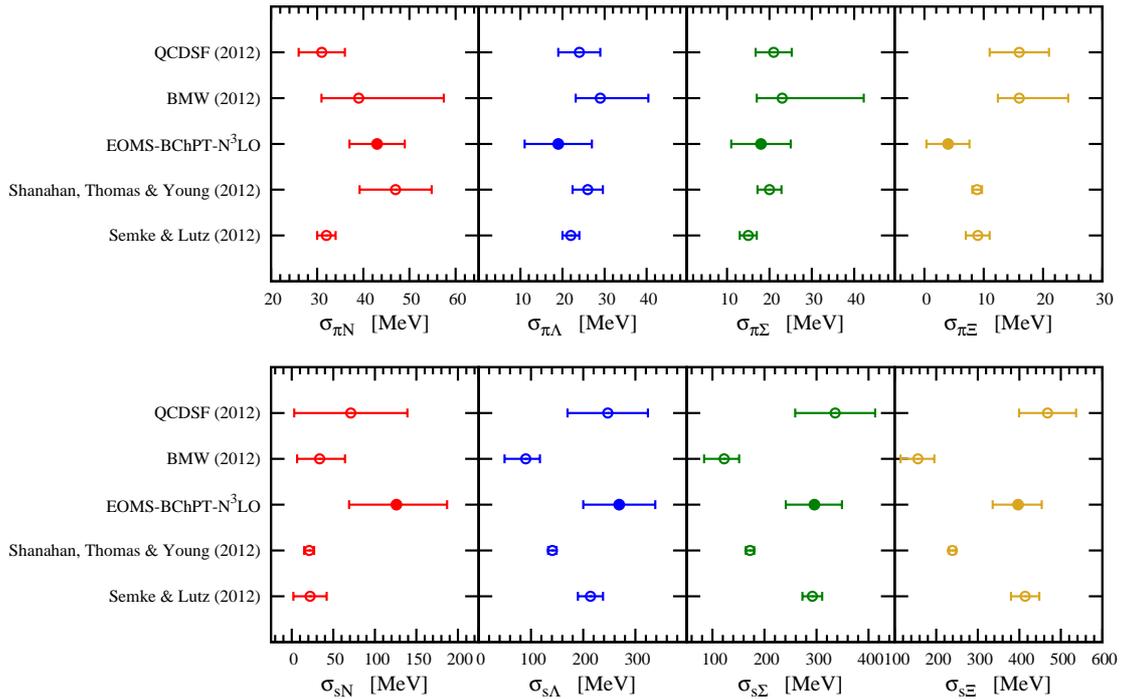}\\
  \caption{(Color online) Pion and strangeness sigma terms of the octet baryons. The solid circles represent our results. The empty circles are the results from the BMW~\cite{Durritetal.(BMWCollaboartion)2012_Phys.Rev.D85_014509}, QCDSF~\cite{Horsleyitetal.(QCDSFCollaboration)2012_Phys.Rev.D85_034506} collaborations and from the BChPT studies with the FRR scheme ~\cite{Shanahan2012__} and the partial summation scheme~\cite{Semke2012_Phys.Lett.B717_242}.}
  \label{sigmaBpis}
\end{figure}

In Fig.~\ref{sigmaBpis}, we compare our results of the sigma terms with the $n_f=2+1$ LQCD results of Refs.~\cite{Durritetal.(BMWCollaboartion)2012_Phys.Rev.D85_014509,Horsleyitetal.(QCDSFCollaboration)2012_Phys.Rev.D85_034506} and the latest results obtained in the  NNLO HBChPT with the FRR scheme \cite{Shanahan2012__} and the N$^3$LO covariant BChPT with the partial summation scheme \cite{Semke2012_Phys.Lett.B717_242}. Our calculated baryon sigma terms are consistent with those of others, except for $\sigma_{\pi\Xi}$ and $\sigma_{sN}$: Our $\sigma_{\pi\Xi}$ is a bit smaller  and $\sigma_{s N}=126(24)(54)$ is slightly larger than the other results,.

The nucleon pion-sigma term at the physical point, $\sigma_{\pi N}=43(1)(6)$ MeV, is in reasonable agreement with the determination in the study of the old $\pi-N$ scattering data \cite{Gasser1991_Phys.Lett.B253_252} of $\sigma_{\pi N}=45\pm 8$ MeV, but smaller than the central value of the more recent study,  $\sigma_{\pi N}=59\pm 7$ MeV~\cite{Alarcon:2011zs}. Our $\sigma_{\pi N}$ is  also in agreement with the recent lattice result of the QCDSF collaboration ($\sigma_{\pi N}=38(12)$ MeV~\cite{Baliitetal.(QCDSFCollaboration)2012_Phys.Rev.D85_054502}, $\sigma_{\pi N}=37(8)(6)$ MeV~\cite{Baliitetal.(QCDSFCollabortion)2013_Nucl.Phys.B866_1}) and BMW Collaboration ($\sigma_{\pi N}=39(4)^{+18}_{-7}$~MeV)~\cite{Durritetal.(BMWCollaboartion)2012_Phys.Rev.D85_014509} within uncertainties, and is consistent with the HBChPT result of Ref.~\cite{Shanahan2012__}. On the other hand, our value is larger than the QCDSF result ($\sigma_{\pi N}=31(3)(4)$ MeV)~\cite{Horsleyitetal.(QCDSFCollaboration)2012_Phys.Rev.D85_034506} and the N$^3$LO BChPT results using the partial summation scheme ($\sigma_{\pi N}=32(1)$ MeV)~\cite{Semke2012_Phys.Lett.B717_242}, but slightly smaller than the JLQCD result ($\sigma_{\pi N}=50(4.5)$ MeV)~\cite{Ohkiitetal.(JLQCDCollaboration)2009_ProceedingsofScienceLAT2009_124}.

It should be mentioned that the central value of the nucleon sigma term $\sigma_{\pi N}$ in our present calculation is smaller than that of Ref.~\cite{MartinCamalich2010_Phys.Rev.D82_074504},
where the result is obtained with NNLO EOMS-BChPT by fitting the PACS-CS data. Inclusion of virtual decuplet contributions may have some non-negligible effects on the predicted baryon sigma terms. At NNLO, it is found that the inclusion of virtual decuplet baryons can increase the pion-nucleon sigma term while decrease the strangeness-nucleon sigma term. It is interesting to check whether such effects still exist at N$^3$LO.

\section{Summary }\label{sec6}

We have studied the lowest-lying octet baryon masses with the  EOMS BChPT  up to N$^3$LO. The unknown low-energy constants are determined by a simultaneous fit of the latest $2+1$ flavor LQCD simulations  from the PACS-CS, LHPC, HSC, QCDSF-UKQCD and NPLQCD collaborations. Finite-volume corrections are calculated self-consistently. It is shown that the eleven lattice data sets with  $M_{\pi}^2<0.25$ GeV$^2$ and $M_{\phi}L>4~(\phi=\pi,~K,~\eta)$ can be fitted with a $\chi^2$/d.o.f.~=~1.0.  Including more lattice data with larger pion masses or smaller volumes deteriorates the fit a bit but still yields a reasonable $\chi^2/\mathrm{d.o.f}=1.6$.

Our studies confirm that covariant BChPT in the three flavor sector converges as expected, i.e., relatively slowly as dictated by $M_K/\Lambda_{\chi\mathrm{SB}}$ but with clear improvement order by order, at least concerning the octet baryon masses. A successful simultaneous fit of all the latest 2+1 flavor LQCD simulations indicates that the LQCD results are consistent with each other, though their setups are quite different.

Applying the Feynman-Hellmann theorem, we  have  calculated the sigma terms of the octet baryons.  In comparison with the results
obtained in previous studies, we conclude that to obtain the sigma terms with an accuracy of  a few percent, more works are still needed.

\acknowledgments

This work was partly supported by the National
Natural Science Foundation of China under Grants No. 11005007, No. 11035007, and No. 11175002,  the New Century Excellent Talents in University  Program of Ministry of Education of China under Grant No. NCET-10-0029,  the Fundamental Research Funds for the Central Universities, and
the Research Fund  for the Doctoral Program of Higher Education under Grant No. 20110001110087.
JMC acknowledges the Spanish Government and FEDER funds under contract FIS2011-28853-C02-01 and the STFC [Grant No. ST/H004661/1] for support.

\appendix

\section{$N_f=2+1$~Lattice QCD simulation results}\label{latticedata}
We briefly summarize some key ingredients of the LQCD simulations of the PACS-CS~\cite{Aokiitetal.(PACS-CSCollaboration)2009_Phys.Rev.D79_034503}, LHPC\cite{Walker-Louditelal.2009_Phys.Rev.D79_054502}, HSC~\cite{Linitetal.(HSCCollaboration)2009_Phys.Rev.D79_034502}, QCDSF-UKQCD\cite{Bietenholzitetal.(QCDSF-UKQCDCollaboration)2011_PhysicalReviewD84_054509} and NPLQCD~\cite{Beaneitetal.(NPLQCDCollaboration)2011_PhysicalReviewD84_014507} collaborations, which are relevant to
our study. In addition, we tabulate the simulated octet baryon masses in physical units, which satisfy $M_{\pi}^2<0.5$ GeV$^2$,  $M_K^2<0.7$ GeV$^2$, and $M_{\phi}L>3$.

\subsection*{PACS-CS ~\cite{Aokiitetal.(PACS-CSCollaboration)2009_Phys.Rev.D79_034503}}

The PACS-CS  collaboration employs the nonperturbatively $O(a)$-improved Wilson quark action and the Iwasaki gauge action. Numerical simulations are carried out at the lattice spacing of $a=0.0907(14)$ fm, on a $32^3\times 64$ lattice with the use of the domain-decomposed HMC algorithm to reduce the up-down quark mass, which is about 3 MeV. For the strange quark part they improve the PHMC algorithm with the UV-filtering procedure. Their simulation points cover from 701 MeV to 156 MeV, but the lightest point has a $M_\pi L\approx2.9$, which might induce large finite volume corrections.
\begin{table}[h!]
\centering
\caption{Masses of the pseudoscalar mesons and the  octet baryons (in units of MeV) obtained by the PACS-CS collaboration (TABLE III of Ref.~\cite{Aokiitetal.(PACS-CSCollaboration)2009_Phys.Rev.D79_034503}.)
The first  number in the parentheses is the statistical uncertainty and second is that propagated from the determination of the lattice spacing.}
\label{PACS-CSdata}
~~\\[0.1em]
\begin{tabular}{rr|cccccc|c}
\hline\hline
  $M_{\pi}$  & $M_K$  & $m_N$        & $m_{\Lambda}$ & $m_{\Sigma}$ & $m_{\Xi}$    \\
\hline
  155.8      &  553.7 & 932.1(78.3)(14.4)   & 1139.9(20.7)(17.6)  & 1218.4(21.5)(18.8) & 1393.3(6.7)(21.5)  \\
  $*$~295.7      &  593.5 & 1093.1(18.9)(16.9) & 1253.8(14.1)(19.4)  & 1314.8(15.4)(20.3) & 1447.7(10.0)(22.3) \\
  $*$~384.4      &  581.4 & 1159.7(15.4)(17.9) & 1274.1(9.1)(19.7)   & 1316.5(10.4)(20.3) & 1408.3(7.0)(21.7)   \\
  $*$~411.2      &  635.0 & 1214.7(11.5)(18.7) & 1350.4(7.8)(20.8)   & 1400.2(8.5)(21.6)  & 1503.1(6.5)(23.2)   \\
  569.7      &  713.2 & 1411.1(12.2)(21.8) & 1503.8(9.8)(23.2)   & 1531.2(11.1)(23.6) & 1609.5(9.4)(24.8)   \\
  701.4      &  789.0 & 1583.0(4.8)(24.4)  & 1643.9(5.0)(25.4)   & 1654.5(4.4)(25.5)  & 1709.6(5.4)(26.4)    \\
  \hline\hline
\end{tabular}
\end{table}

\subsection*{LHPC~\cite{Walker-Louditelal.2009_Phys.Rev.D79_054502}}
The LHPC collaboration calculates the light hadron spectrum in full QCD using a mixed action that exploits the lattice chiral symmetry provided by domain wall valence quarks (the DWF valence quark action) and ensembles of computationally economical improved staggered sea quark configurations~(the so-called asqtad action). The lattice spacing is determined to be $a=0.12406(248)$~fm and the lattice volume is $20^3\times 64$ . The range of pion masses simulated in this work extends from $758$ MeV down to $293$ MeV.

\begin{table}[h!]
\centering
\caption{Masses of the pseudoscalar mesons and the octet baryons (in units of MeV)  obtained by the LHPC collaboration (TABLE II, TABLE VI, TABLE VII of Ref.~\cite{Walker-Louditelal.2009_Phys.Rev.D79_054502}).
The uncertainties have the same origin as those in Table~\ref{PACS-CSdata}.}
\label{LHPCdata}
~~\\[0.1em]
\begin{tabular}{rr|cccccc|c}
\hline\hline
  $M_{\pi}$  & $M_K$  & $m_N$        & $m_{\Lambda}$ & $m_{\Sigma}$ & $m_{\Xi}$    \\
\hline
  292.9      &  585.6 & 1098.9(8.0)(22.0)  & 1240.5(4.8)(24.8)   & 1321.6(6.4)(26.4)  & 1412.2(3.2)(28.2)  \\
  $*$~355.9      &  602.9 & 1157.8(6.4)(23.1)  & 1280.2(4.8)(25.6)   & 1350.2(4.8)(27.0)  & 1432.9(3.2)(28.6)  \\
  $*$~495.1      &  645.4 & 1288.2(6.4)(25.8)  & 1369.3(4.8)(27.4)   & 1409.1(6.4)(28.2)  & 1469.5(4.8)(29.4)  \\
  596.7      &  685.6 & 1394.8(6.4)(27.9)  & 1440.9(8.0)(28.8)   & 1463.1(9.5)(29.2)  & 1504.5(8.0)(30.1)  \\
  687.7      &  728.1 & 1502.9(11.1)(30.0) & 1528.3(9.5)(30.6)   & 1536.3(9.5)(30.7)  & 1557.0(9.5)(31.1)  \\
  \hline\hline
\end{tabular}
\end{table}

\subsection*{HSC~\cite{Linitetal.(HSCCollaboration)2009_Phys.Rev.D79_034502}}
The HSC collaboration uses a Symanzik-improved action with tree-level tadpole-improved coefficients for the gauge sector
 and  the anisotropic clover fermion action for the fermion sector.  The lattice spacings are  $a_s=0.1227(8)$ fm and $a_t=0.003506(23)$ fm in spatial and temporal directions, respectively. The simulations are performed at
  four different lattice volumes $12^3\times96$, $16^3\times96$, $16^3\times128$, and  $24^3\times128$. The simulated pion masses range
   from 383 MeV to 1565 MeV. For our purposes, we only need those data with $M_{\pi}^2\leq 0.5$ GeV$^2$. The pseudoscalar meson masses and corresponding octet baryon masses are listed in Table \ref{HSCdata}.

\begin{table}[h!]
\centering
\caption{Masses of the pseudoscalar mesons and the lowest-lying baryons (in units of MeV) obtained by the HSC collaboration (TABLE VI and TABLE VII of Ref.~\cite{Linitetal.(HSCCollaboration)2009_Phys.Rev.D79_034502}). The uncertainties have the same origin as those in Table~\ref{PACS-CSdata}.}
\label{HSCdata}
~~\\[0.1em]
\begin{tabular}{rr|cccccc|c}
\hline\hline
  $M_{\pi}$  & $M_K$  & $m_N$        & $m_{\Lambda}$ & $m_{\Sigma}$ & $m_{\Xi}$    \\
\hline
  $*$~383.2      &  543.6 & 1147.5(10.7)(7.5)   & 1243.1(8.4)(8.2)     & 1287.0(8.4)(8.4)     & 1347.8(6.8)(8.8)           \\
  388.9      &  545.9 & 1164.9(22.5)(7.6)   & 1226.8(16.9)(8.0)   & 1288.7(16.9)(8.5)   & 1345.0(11.3)(8.8)        \\
  $*$~448.5      &  580.8 & 1238.1(16.9)(8.1)   & 1328.1(11.3)(8.7)   & 1361.9(16.9)(8.9)   & 1412.5(10.7)(9.3)        \\
  560.5      &  646.6 & 1361.9(22.5)(8.9)   & 1440.6(16.9)(9.5)   & 1457.5(22.5)(9.6)   & 1496.9(16.9)(9.8)        \\
  \hline\hline
\end{tabular}
\end{table}

\subsection*{QCDSF-UKQCD~\cite{Bietenholzitetal.(QCDSF-UKQCDCollaboration)2011_PhysicalReviewD84_054509}}
The QCDSF-UKQCD collaboration employs the particular clover action, which has a single iterated mild stout smearing, and the (tree-level) Symanzik improved gluon action, which contains the gluon action and the three-flavor Wilson-Dirac fermion action.
The simulations are carried out at the lattice spacing of $a\sim 0.075-0.078$ fm, on $16^3\times 32$, $24^3\times 48$ and $32^3\times 64$ lattices.  The resulting pion masses range from $229$ MeV to $449$ MeV. In the simulations, they kept the singlet quark mass fixed and tuned the quark masses to ensure that the kaon always has a mass less than the physical one.  It should be noted that in Table \ref{UKQCDdata}
 we did not tabulate the $16^3\times 32$ and the $32^3\times 64$ three-flavor simulation results, which have meson masses out of the range specified above.

\begin{table}[h!]
\centering
\caption{Masses of the pseudoscalar mesons and the lowest-lying baryons (in units of MeV) obtained by the QCDSF-UKQCD collaboration (TABLE XX, TABLE XXII and TABLE XXIII of Ref.~\cite{Bietenholzitetal.(QCDSF-UKQCDCollaboration)2011_PhysicalReviewD84_054509}). The uncertainties have the same origin as those in Table~\ref{PACS-CSdata}.}
\label{UKQCDdata}
~~\\[0.1em]
\begin{tabular}{rr|cccccc|c}
\hline\hline
  $M_{\pi}$  & $M_K$  & $m_N$        & $m_{\Lambda}$ & $m_{\Sigma}$ & $m_{\Xi}$    \\ 
\hline
  228.9      &  476.0 & 999.2(13.6)(3.8)  & 1144.2(15.0)(4.3)  & 1184.6(5.8)(4.5)  & 1266.6(9.2)(4.8) \\ 
  275.2      &  463.2 & 1035.0(8.9)(3.9)  & 1127.6(12.8)(4.3)  & 1172.0(4.6)(4.4)  & 1243.3(9.2)(4.7) \\
  $*$~320.4      &  450.6 & 1061.4(5.5)(4.0)  & 1130.9(6.7)(4.3)   & 1168.5(2.3)(4.4)  & 1220.3(4.6)(4.6)  \\
  324.2      &  449.5 & 1071.8(6.4)(4.0)  & 1154.7(8.1)(4.4)   & 1162.8(2.3)(4.4)  & 1216.8(4.6)(4.6)  \\
  350.7      &  439.4 & 1097.5(3.7)(4.1)  & 1136.9(8.9)(4.3)   & 1159.3(2.3)(4.4)  & 1193.8(3.5)(4.5)  \\
  383.0      &  425.5 & 1123.5(3.8)(4.2)  & 1137.1(9.7)(4.3)   & 1155.9(1.2)(4.4)  & 1170.8(3.4)(4.4)  \\
  411.5      &  411.5 & 1150.1(3.8)(4.3)  & 1150.1(3.8)(4.3)   & 1150.1(3.8)(4.3)  & 1150.1(3.8)(4.3)  \\
  448.8      &  391.9 & 1188.1(2.3)(4.5)  & 1160.5(6.9)(4.4)   & 1144.2(1.5)(4.3)  & 1117.6(3.0)(4.2) \\
\hline\hline
\end{tabular}
\end{table}

\subsection*{NPLQCD~\cite{Beaneitetal.(NPLQCDCollaboration)2011_PhysicalReviewD84_014507}}
The NPLQCD collaboration mainly studied  finite-volume effects on the octet baryon masses. Simulations are performed with $n_f=2+1$ anisotropic clover Wilson action in four lattice volumes with spatial extent $L\sim 2.0,~2.5,~3.0$ and $3.9$ fm. The anisotropic lattice spacing in the spatial direction is $b_s\sim 0.123$ fm and $b_t=b_s/3.5$ in the time direction.  The pion mass is fixed at $M_{\pi}\sim 390$ MeV.

\begin{table}[h!]
\centering
\caption{Masses of the pseudoscalar mesons and the lowest-lying baryons (in units of MeV) obtained by the NPLQCD collaboration (TABLE II of Ref.~\cite{Beaneitetal.(NPLQCDCollaboration)2011_PhysicalReviewD84_014507}). The uncertainties have the same origin as those in Table~\ref{PACS-CSdata}.}
\label{NPLQCDdata}
~~\\[0.1em]
\begin{tabular}{rr|cccccc|c}
\hline\hline
  $M_{\pi}$  & $M_K$  & $m_N$        & $m_{\Lambda}$ & $m_{\Sigma}$ & $m_{\Xi}$    \\
\hline
387.8  &  544.4 & 1182.1(5.4)(7.7) & 1263.3(5.1)(8.2) & 1286.6(4.3)(8.4) & 1361.5(4.1)(8.9) \\
$*$~387.8  &  544.4 & 1164.0(3.2)(7.6) & 1252.0(2.6)(8.2) & 1280.5(3.0)(8.3) & 1356.4(2.6)(8.8) \\
$*$~387.8  &  544.4 & 1151.6(2.5)(7.5) & 1242.3(2.6)(8.1) & 1282.7(2.2)(8.4) & 1349.3(2.1)(8.8) \\
$*$~387.8  &  544.4 & 1151.3(2.6)(7.5) & 1241.2(2.2)(8.1) & 1279.0(2.8)(8.3) & 1349.2(2.0)(8.9) \\
\hline\hline
\end{tabular}
\end{table}

\section{Linear fits of $M_{K}^2 = a +b M_{\pi}^2$}\label{app2}
\begin{figure}[t]
\centering
  \includegraphics[width=12cm]{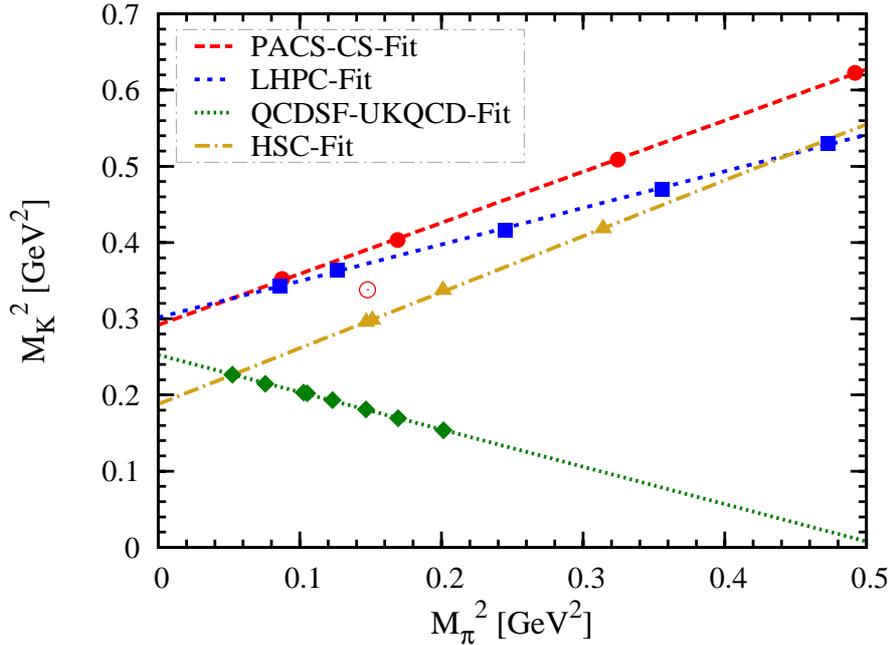}\\
  \caption{(Color online) $M_{K}^2$ vs. $M_{\pi}^2$. The dashed line is the fit to the PACS-CS data, hashed line the fit to the LHPC data, dotted line the fit to the QCDSF-UKQCD data, and dot-dashed line the fit to the HSC  data. In the fit of the PACS-CS data,
  the ensemble with a lower strange-quark mass (the red hollow circle) is not included. }
\label{mkmpilinear}
\end{figure}

In this section, we determine the coefficients of the linear functions, $M_{K}^2 = a +b M_{\pi}^2$, by fitting the lattice data from the PACS-CS, LHPC, QCDSF-UKQCD and HSC collaborations. The fitting results are displayed in Fig.~\ref{mkmpilinear}. It should be noted that the data point of
the PACS-CS collaboration denoted by a red hollow circle is not included in the corresponding fit because it has a lower strange-quark mass. The
explicit expressions of the fitting functions are:
\begin{itemize}
\item PACS-CS:
\begin{equation}
  M_K^2=0.291751 + 0.670652 M_{\pi}^2.
\end{equation}
\item LHPC:
\begin{equation}
  M_K^2=0.301239 + 0.479545 M_{\pi}^2.
\end{equation}
\item QCDSF-UKQCD:
\begin{equation}
  M_K^2=0.252658 - 0.489594 M_{\pi}^2.
\end{equation}
\item HSC:
\begin{equation}
  M_K^2=0.187873 + 0.734493 M_{\pi}^2.
\end{equation}
\end{itemize}

\newpage

\bibliographystyle{JHEP}
\bibliography{NNLO}

\providecommand{\href}[2]{#2}\begingroup\raggedright\begin{thebibliography}{10}

\bibitem{Wilson1974_Phys.Rev.D10_2445}
K.~G. Wilson, {\it {Confinement of quarks}},  {\em Phys. Rev. D} {\bf 10}
  (1974) 2445.

\bibitem{Gattringer2010__}
C.~Gattringer and C.~B. Lang, {\em Quantum Chromodynamics on the Lattice - An
  Introductory Presentation}.
\newblock Springer, Heidelberg, 2010.

\bibitem{Alexandrouitetal.2009_Phys.Rev.D80_114503}
C.~Alexandrou~{\it et al.}, {\it {Low-lying baryon spectrum with two dynamical
  twisted mass fermions}},  {\em Phys. Rev. D} {\bf 80} (2009) 114503,
  [\href{http://xxx.lanl.gov/abs/0910.2419}{{\tt arXiv:0910.2419}}].

\bibitem{Durritetal.2009_Science322_1224}
S.~D\"{u}rr~{\it et al.}, {\it {Ab-initio Determination of Light Hadron
  Masses}},  {\em Science} {\bf 322} (2009) 1224,
  [\href{http://xxx.lanl.gov/abs/0906.3599}{{\tt arXiv:0906.3599}}].

\bibitem{Aokiitetal.(PACS-CSCollaboration)2009_Phys.Rev.D79_034503}
S.~Aoki~{\it et al.} (PACS-CS~Collaboration), {\it {2+1 flavor lattice QCD
  toward the physical point}},  {\em Phys. Rev. D} {\bf 79} (2009) 034503,
  [\href{http://xxx.lanl.gov/abs/0807.1661}{{\tt arXiv:0807.1661}}].

\bibitem{Aokiitetal.(PACS-CSCollaboration)2010_Phys.Rev.D81_074503}
S.~Aoki~{\it et al.} (PACS-CS~Collaboration), {\it {Physical point simulation
  in 2+1 flavor lattice QCD}},  {\em Phys. Rev. D} {\bf 81} (2010) 074503,
  [\href{http://xxx.lanl.gov/abs/0911.2561}{{\tt arXiv:0911.2561}}].

\bibitem{Walker-Louditelal.2009_Phys.Rev.D79_054502}
A.~Walker-Loud~{\it el al.}, {\it {Light hadron spectroscopy using domain wall
  valence quarks on an Asqtad sea}},  {\em Phys. Rev. D} {\bf 79} (2009)
  054502, [\href{http://xxx.lanl.gov/abs/0806.4549}{{\tt arXiv:0806.4549}}].

\bibitem{Linitetal.(HSCCollaboration)2009_Phys.Rev.D79_034502}
H.-W. Lin~{\it et al.} (HSC~Collaboration), {\it {First results from 2+1
  dynamical quark flavors on an anisotropic lattice: light-hadron spectroscopy
  and setting the strange-quark mass}},  {\em Phys. Rev. D} {\bf 79} (2009)
  034502, [\href{http://xxx.lanl.gov/abs/0810.3588}{{\tt arXiv:0810.3588}}].

\bibitem{Bietenholzitetal.2010_Phys.Lett.B690_436--441}
W.~Bietenholz~{\it et al.}, {\it {Tuning the strange quark mass in lattice
  simulations}},  {\em Phys. Lett. B} {\bf 690} (2010) 436,
  [\href{http://xxx.lanl.gov/abs/1003.1114}{{\tt arXiv:1003.1114}}].

\bibitem{Bietenholzitetal.(QCDSF-UKQCDCollaboration)2011_PhysicalReviewD84_054509}
W.~Bietenholz~{\it et al.} (QCDSF-UKQCD~Collaboration), {\it {Flavor blindness
  and patterns of flavor symmetry breaking in lattice simulations of up, down,
  and strange quarks}},  {\em Phys. Rev. D} {\bf 84} (2011) 054509.

\bibitem{Beaneitetal.(NPLQCDCollaboration)2011_PhysicalReviewD84_014507}
S.~R. Beane~{\it et al.} (NPLQCD~Collaboration), {\it {High statistics analysis
  using anisotropic clover lattices: IV. Volume dependence of light hadron
  masses}},  {\em Phys. Rev. D} {\bf 84} (2011) 014507,
  [\href{http://xxx.lanl.gov/abs/1104.4101}{{\tt arXiv:1104.4101}}].

\bibitem{Beringeritetal.(ParticleDataGroup)2012_Phys.Rev.D86_010001}
J.~Beringer~{\it et al.} (Particle Data~Group), {\it {The Review of Particle
  Physics}},  {\em Phys. Rev. D} {\bf 86} (2012) 010001.

\bibitem{Fodor2012_Rev.Mod.Phys.84_449}
Z.~Fodor and C.~Hoelbling, {\it {Light Hadron Masses from Lattice QCD}},  {\em
  Rev. Mod. Phys.} {\bf 84} (2012) 449,
  [\href{http://xxx.lanl.gov/abs/1203.4789}{{\tt arXiv:1203.4789}}].

\bibitem{Weinberg1979_PhysicaA96_327--340}
S.~Weinberg, {\it {Phenomenological Lagrangians}},  {\em Physica A} {\bf 96}
  (1979) 327.

\bibitem{Gasser1984_Ann.Phys.(N.Y.)158_142--210}
J.~Gasser and H.~Leutwyler, {\it {Chiral perturbation theory to one loop}},
  {\em Ann. Phys. (N.Y.)} {\bf 158} (1984) 142.

\bibitem{Gasser1985_Nucl.Phys.B250_465--516}
J.~Gasser and H.~Leutwyler, {\it {Chiral perturbation theory: Expansions in the
  mass of the strange quark}},  {\em Nucl. Phys. B} {\bf 250} (1985) 465.

\bibitem{Gasser1988_Nucl.Phys.B307_779--853}
J.~Gasser, M.~E. Sainio, and A.~Svarc, {\it {Nucleons with chiral loops}},
  {\em Nucl. Phys. B} {\bf 307} (1988) 779.

\bibitem{Leutwyler1994_Lecture_}
H.~Leutwyler, {\it {Principles of Chiral Perturbation Theory}},  {\em Lecture}
  (1994) [\href{http://xxx.lanl.gov/abs/hep-ph/9406283}{{\tt hep-ph/9406283}}].

\bibitem{Bernard1995_Int.J.Mod.Phys.E4_193--346}
V.~Bernard, N.~Kaiser, and U.-G. Mei{\ss}ner, {\it {Chiral Dynamics in Nucleons
  and Nuclei}},  {\em Int. J. Mod. Phys. E} {\bf 4} (1995) 193,
  [\href{http://xxx.lanl.gov/abs/hep-ph/9501384}{{\tt hep-ph/9501384}}].

\bibitem{Pich1995_Rep.Prog.Phys.58_57}
A.~Pich, {\it {Chiral Perturbation Theory}},  {\em Rep. Prog. Phys.} {\bf 58}
  (1995) 57, [\href{http://xxx.lanl.gov/abs/hep-ph/9502366}{{\tt
  hep-ph/9502366}}].

\bibitem{Ecker1995_Prog.Part.Nucl.Phys.35_84}
G.~Ecker, {\it {Chiral perturbation theory}},  {\em Prog. Part. Nucl. Phys.}
  {\bf 35} (1995) 84, [\href{http://xxx.lanl.gov/abs/hep-ph/9501357}{{\tt
  hep-ph/9501357}}].

\bibitem{Pich1998_arXiv_}
A.~Pich, {\it {Effective Field Theory}},
  \href{http://xxx.lanl.gov/abs/hep-ph/9806303}{{\tt hep-ph/9806303}}.

\bibitem{Bernard2007_Annu.Rev.Nucl.Part.Sci.57_33--60}
V.~Bernard and U.-G. Mei{\ss}ner, {\it {Chiral Perturbation Theory}},  {\em
  Annu. Rev. Nucl. Part. Sci.} {\bf 57} (2007) 33,
  [\href{http://xxx.lanl.gov/abs/hep-ph/0611231}{{\tt hep-ph/0611231}}].

\bibitem{Bernard2008_Prog.Part.Nucl.Phys.60_82--160}
V.~Bernard, {\it {Chiral perturbation theory and baryon properties}},  {\em
  Prog. Part. Nucl. Phys.} {\bf 60} (2008) 82.

\bibitem{Scherer2012__}
S.~Scherer and M.~R. Schindler, {\em {A Primer for Chiral Perturbation
  Theory}}.
\newblock Springer, Heidelberg, 2012.

\bibitem{Jenkins1991_Phys.Lett.B255_558--562}
E.~E. Jenkins and A.~V. Manohar, {\it {Baryon chiral perturbation theory using
  a heavy fermion lagrangian}},  {\em Phys. Lett. B} {\bf 255} (1991) 558.

\bibitem{Becher1999_Eur.Phys.J.C9_643--671}
T.~Becher and H.~Leutwyler, {\it {Baryon chiral perturbation theory in
  manifestly Lorentz invariant form}},  {\em Eur. Phys. J. C} {\bf 9} (1999)
  643.

\bibitem{Gegelia1999_Phys.Rev.D60_114038}
J.~Gegelia and G.~Japaridze, {\it {Matching Heavy Particle Approach to
  Relativistic Theory}},  {\em Phys. Rev. D} {\bf 60} (1999) 114038,
  [\href{http://xxx.lanl.gov/abs/hep-ph/9908377}{{\tt hep-ph/9908377}}].

\bibitem{Fuchs2003_Phys.Rev.D68_056005}
T.~Fuchs, J.~Gegelia, G.~Japaridze, and S.~Scherer, {\it {Renormalization of
  relativistic baryon chiral perturbation theory and power counting}},  {\em
  Phys. Rev. D} {\bf 68} (2003) 056005,
  [\href{http://xxx.lanl.gov/abs/hep-ph/0302117}{{\tt hep-ph/0302117}}].

\bibitem{Bernard2004_Nucl.Phys.A732_149--170}
V.~Bernard, T.~R. Hemmert, and U.-G. Mei{\ss}ner, {\it {Cutoff schemes in
  chiral perturbation theory and the quark mass expansion of the nucleon
  mass}},  {\em Nucl. Phys. A} {\bf 732} (2004) 149,
  [\href{http://xxx.lanl.gov/abs/hep-ph/0307115}{{\tt hep-ph/0307115}}].

\bibitem{Young2003_Prog.Part.Nucl.Phys.50_399--417}
R.~D. Young, D.~B. Leinweber, and A.~W. Thomas, {\it {Convergence of Chiral
  Effective Field Theory Chiral Effective}},  {\em Prog. Part. Nucl. Phys.}
  {\bf 50} (2003) 399.

\bibitem{Leinweber2003_Phys.Rev.Lett.92_24}
D.~B. Leinweber, A.~W. Thomas, and R.~D. Young, {\it {Physical Nucleon
  Properties from Lattice QCD}},  {\em Phys. Rev. Lett.} {\bf 92} (2003) 24,
  [\href{http://xxx.lanl.gov/abs/hep-lat/0302020}{{\tt hep-lat/0302020}}].

\bibitem{Young2010_Phys.Rev.D81_014503}
R.~D. Young and A.~W. Thomas, {\it {Octet baryon masses and sigma terms from an
  SU(3) chiral extrapolation}},  {\em Phys. Rev. D} {\bf 81} (2010) 014503,
  [\href{http://xxx.lanl.gov/abs/0901.3310}{{\tt arXiv:0901.3310}}].

\bibitem{Semke2006_Nucl.Phys.A778_153--180}
A.~Semke and M.~F.~M. Lutz, {\it {Baryon self energies in the chiral loop
  expansion}},  {\em Nucl. Phys. A} {\bf 778} (2006) 153,
  [\href{http://xxx.lanl.gov/abs/nucl-th/0511061}{{\tt nucl-th/0511061}}].

\bibitem{Jenkins1992_Nucl.Phys.B368_190}
E.~E. Jenkins, {\it {Baryon masses in chiral perturbation theory}},  {\em Nucl.
  Phys. B} {\bf 368} (1992) 190.

\bibitem{Bernard1993_Z.PhysikC60_111--119}
V.~Bernard, N.~Kaiser, and U.-G. Mei{\ss}ner, {\it {Critical Analysis of Baryon
  Masses and Sigma-Terms in Heavy Baryon Chiral Perturbation Theory}},  {\em Z.
  Physik C} {\bf 60} (1993) 111,
  [\href{http://xxx.lanl.gov/abs/hep-ph/9303311}{{\tt hep-ph/9303311}}].

\bibitem{Banerjee1995_Phys.Rev.D52_11}
M.~K. Banerjee and J.~Milana, {\it {Baryon mass splittings in chiral
  perturbation theory}},  {\em Phys. Rev. D} {\bf 52} (1995) 11.

\bibitem{Borasoy1996_Ann.Phys.(N.Y.)254_192--232}
B.~Borasoy and U.-G. Mei{\ss}ner, {\it {Chiral expansion of baryon masses and
  $\sigma$-terms}},  {\em Ann. Phys. (N.Y.)} {\bf 254} (1996) 192,
  [\href{http://xxx.lanl.gov/abs/hep-ph/9607432}{{\tt hep-ph/9607432}}].

\bibitem{Walker-Loud2004_Nucl.Phys.A747_476--507}
A.~Walker-Loud, {\it {Octet Baryon Masses in Partially Quenched Chiral
  Perturbation Theory}},  {\em Nucl. Phys. A} {\bf 747} (2004) 476,
  [\href{http://xxx.lanl.gov/abs/hep-lat/0405007}{{\tt hep-lat/0405007}}].

\bibitem{Ellis1999_Phys.Rev.C61_015205}
P.~J. Ellis and K.~Torikoshi, {\it {Baryon Masses in Chiral Perturbation Theory
  with Infrared Regularization}},  {\em Phys. Rev. C} {\bf 61} (1999) 015205,
  [\href{http://xxx.lanl.gov/abs/nucl-th/9904017}{{\tt nucl-th/9904017}}].

\bibitem{Frink2004_JHEP07_028}
M.~Frink and U.-G. Mei{\ss}ner, {\it {Chiral extrapolations of baryon masses
  for unquenched three-flavor lattice simulations}},  {\em JHEP} {\bf 07}
  (2004) 028, [\href{http://xxx.lanl.gov/abs/hep-lat/0404018}{{\tt
  hep-lat/0404018}}].

\bibitem{Frink2005_Eur.Phys.J.A.24_395}
M.~Frink, U.-G. Mei{\ss}ner, and I.~Scheller, {\it {Baryon masses, chiral
  extrapolations, and all that}},  {\em Eur. Phys. J. A.} {\bf 24} (2005) 395,
  [\href{http://xxx.lanl.gov/abs/hep-lat/0501024}{{\tt hep-lat/0501024}}].

\bibitem{Lehnhart2004_J.Phys.G:Nucl.Part.Phys.31_89}
B.~C. Lehnhart, J.~Gegelia, and S.~Scherer, {\it {Baryon masses and nucleon
  sigma terms in manifestly Lorentz-invariant baryon chiral perturbation
  theory}},  {\em J. Phys. G: Nucl. Part. Phys.} {\bf 31} (2004) 89,
  [\href{http://xxx.lanl.gov/abs/hep-ph/0412092}{{\tt hep-ph/0412092}}].

\bibitem{MartinCamalich2010_Phys.Rev.D82_074504}
J.~{Martin Camalich}, L.~S. Geng, and M.~J. {Vicente Vacas}, {\it {Lowest-lying
  baryon masses in covariant SU(3)-flavor chiral perturbation theory}},  {\em
  Phys. Rev. D} {\bf 82} (2010) 074504,
  [\href{http://xxx.lanl.gov/abs/1003.1929}{{\tt arXiv:1003.1929}}].

\bibitem{Semke2007_Nucl.Phys.A789_251--259}
A.~Semke and M.~F.~M. Lutz, {\it {On the possibility of a discontinuous
  quark-mass dependence of baryon octet and decuplet masses}},  {\em Nucl.
  Phys. A} {\bf 789} (2007) 251.

\bibitem{Semke2012_Phys.Rev.D85_034001}
A.~Semke and M.~F.~M. Lutz, {\it {Quark-mass dependence of the baryon
  ground-state masses}},  {\em Phys. Rev. D} {\bf 85} (2012) 034001.

\bibitem{Bruns:2012eh}
P.~C. Bruns, L.~Greil, and A.~Schafer, {\it {Chiral extrapolation of baryon
  mass ratios}},  \href{http://xxx.lanl.gov/abs/1209.0980}{{\tt
  arXiv:1209.0980}}.

\bibitem{Lutz2012_Phys.Rev.D86_091502(R)}
M.~F.~M. Lutz and A.~Semke, {\it {On the consistency of recent QCD lattice data
  of the baryon ground-state masses}},  {\em Phys. Rev. D} {\bf 86} (2012)
  091502(R), [\href{http://xxx.lanl.gov/abs/1209.2791}{{\tt arXiv:1209.2791}}].

\bibitem{Ishikawa2009_Phys.Rev.D80_054502}
K.~I. Ishikawa~{\it et al.} (PACS-CS~Collaboration), {\it {SU(2) and SU(3)
  chiral perturbation theory analyses on baryon masses in 2+1 flavor lattice
  QCD}},  {\em Phys. Rev. D} {\bf 80} (2009) 054502,
  [\href{http://xxx.lanl.gov/abs/0905.0962}{{\tt arXiv:0905.0962}}].

\bibitem{Alarcon:2012nr}
J.~M. Alarcon, L.~S. Geng, J.~Martin~Camalich, and J.~A. Oller, {\it {On the
  strangeness content of the nucleon}},
  \href{http://xxx.lanl.gov/abs/1209.2870}{{\tt arXiv:1209.2870}}.

\bibitem{Oller2006_JHEP09_079}
J.~A. Oller, M.~Verbeni, and J.~Prades, {\it {Meson-baryon effective chiral
  Lagrangians to $\mathcal{O}(q^3)$}},  {\em JHEP} {\bf 09} (2006) 079.

\bibitem{Geng2008_Phys.Rev.Lett.101_222002}
L.~S. Geng, J.~Martin~Camalich, L.~Alvarez-Ruso, and M.~J. Vicente~Vacas, {\it
  {Leading SU(3)-Breaking Corrections to the Baryon Magnetic Moments in Chiral
  Perturbation Theory}},  {\em Phys. Rev. Lett.} {\bf 101} (2008) 222002.

\bibitem{Geng2011_Phys.Lett.B696_390}
L.~S. Geng, M.~Altenbuchinger, and W.~Weise, {\it {Light quark mass dependence
  of the D and decay constants}},  {\em Phys. Lett. B} {\bf 696} (2011) 390.

\bibitem{Geng2010_ChinesePhys.C34_1307}
L.~S. Geng, J.~Martin~Camalich, L.~Alvarez-Ruso, and M.~J.~V. Vacas, {\it
  {Lowest lying spin-1/2 and spin-3/2 baryon magnetic moments in chiral
  perturbation theory*}},  {\em Chinese Phys. C} {\bf 34} (2010) 1307,
  [\href{http://xxx.lanl.gov/abs/1001.0465}{{\tt arXiv:1001.0465}}].

\bibitem{Bijnens:2011tb}
J.~Bijnens and I.~Jemos, {\it {A new global fit of the $L^r_i$ at
  next-to-next-to-leading order in Chiral Perturbation Theory}},  {\em Nucl.
  Phys. B} {\bf 854} (2012) 631--665,
  [\href{http://xxx.lanl.gov/abs/1103.5945}{{\tt arXiv:1103.5945}}].

\bibitem{Khanetal.(QCDSF-UKQCDCollaboration)2003_Nucl.Phys.B689_175}
A.~A. Khan {et al.} (QCDSF-UKQCD~Collaboration), {\it {The nucleon mass in
  $N_f=2$ lattice QCD: finite size effects from chiral perturbation theory}},
  {\em Nucl. Phys. B} {\bf 689} (2003) 175,
  [\href{http://xxx.lanl.gov/abs/hep-lat/0312030}{{\tt hep-lat/0312030}}].

\bibitem{Procura2006_Phys.Rev.D73_114510}
M.~Procura, B.~U. Musch, T.~Wollenweber, T.~R. Hemmert, and W.~Weise, {\it
  {Nucleon mass: from lattice QCD to the chiral limit}},  {\em Phys. Rev. D}
  {\bf 73} (2006) 114510, [\href{http://xxx.lanl.gov/abs/hep-lat/0603001}{{\tt
  hep-lat/0603001}}].

\bibitem{Geng2011_Phys.Rev.D84_074024}
L.~S. Geng, X.-L. Ren, J.~Martin~Camalich, and W.~Weise, {\it {Finite-volume
  effects on octet-baryon masses in covariant baryon chiral perturbation
  theory}},  {\em Phys. Rev. D} {\bf 84} (2011) 074024.

\bibitem{Ohkiitetal.(JLQCDCollaboration)2009_ProceedingsofScienceLAT2009_124}
H.~Ohki~{\it et al.} (JLQCD~Collaboration), {\it {Nucleon sigma term and
  strange quark content in 2+1-flavor QCD with dynamical overlap fermions}},
  {\em Proceedings of Science} {\bf LAT2009} (2009) 124,
  [\href{http://xxx.lanl.gov/abs/0910.3271}{{\tt arXiv:0910.3271}}].

\bibitem{Amoros2001_Nucl.Phys.B602_87}
G.~Amoros, J.~Bijnens, and P.~Talavera, {\it {QCD isospin breaking in meson
  masses, decay constants and quark mass ratios}},  {\em Nucl. Phys. B} {\bf
  602} (2001) 87.

\bibitem{Procura2003_Phys.Rev.D69_034505}
M.~Procura, T.~R. Hemmert, and W.~Weise, {\it {Nucleon mass, sigma term and
  lattice QCD}},  {\em Phys. Rev. D} {\bf 69} (2003) 034505,
  [\href{http://xxx.lanl.gov/abs/hep-lat/0309020}{{\tt hep-lat/0309020}}].

\bibitem{Giedt2009_Phys.Rev.Lett.103_201802}
J.~Giedt, A.~W. Thomas, and R.~D. Young, {\it {Dark matter, the CMSSM and
  lattice QCD}},  {\em Phys. Rev. Lett.} {\bf 103} (2009) 201802,
  [\href{http://xxx.lanl.gov/abs/0907.4177}{{\tt arXiv:0907.4177}}].

\bibitem{Durritetal.(BMWCollaboartion)2012_Phys.Rev.D85_014509}
S.~D\"{u}rr~{\it et al.} (BMW~Collaboartion), {\it {Sigma term and strangeness
  content of octet baryons}},  {\em Phys. Rev. D} {\bf 85} (2012) 014509.

\bibitem{Shanahan2012__}
P.~E. Shanahan, A.~W. Thomas, and R.~D. Young, {\it {Sigma terms from an SU(3)
  chiral extrapolation}},  \href{http://xxx.lanl.gov/abs/1205.5365}{{\tt
  arXiv:1205.5365}}.

\bibitem{Semke2012_Phys.Lett.B717_242}
A.~Semke and M.~F.~M. Lutz, {\it {Strangeness in the baryon ground states}},
  {\em Phys. Lett. B} {\bf 717} (2012) 242,
  [\href{http://xxx.lanl.gov/abs/1202.3556}{{\tt arXiv:1202.3556}}].

\bibitem{Horsleyitetal.(QCDSFCollaboration)2012_Phys.Rev.D85_034506}
R.~Horsley~{\it et al.} (QCDSF~Collaboration), {\it {Hyperon sigma terms for
  2+1 quark flavors}},  {\em Phys. Rev. D} {\bf 85} (2012) 034506.

\bibitem{Gasser1991_Phys.Lett.B253_252}
J.~Gasser, H.~Leutwyler, and M.~E. Sainio, {\it {Sigma term update}},  {\em
  Phys. Lett. B} {\bf 253} (1991) 252.

\bibitem{Alarcon:2011zs}
J.~M. Alarcon, J.~Martin~Camalich, and J.~A. Oller, {\it {The chiral
  representation of the $\pi N$ scattering amplitude and the pion-nucleon sigma
  term}},  {\em Phys.Rev.} {\bf D85} (2012) 051503,
  [\href{http://xxx.lanl.gov/abs/1110.3797}{{\tt arXiv:1110.3797}}].

\bibitem{Baliitetal.(QCDSFCollaboration)2012_Phys.Rev.D85_054502}
G.~Bali~{\it et al.} (QCDSF~Collaboration), {\it {Strange and light quark
  contributions to the nucleon mass from lattice QCD}},  {\em Phys. Rev. D}
  {\bf 85} (2012) 054502.

\bibitem{Baliitetal.(QCDSFCollabortion)2013_Nucl.Phys.B866_1}
G.~Bali~{\it et al.} (QCDSF~Collabortion), {\it {Nucleon mass and sigma term
  from lattice QCD with two light fermion flavors}},  {\em Nucl. Phys. B} {\bf
  866} (2013) 1, [\href{http://xxx.lanl.gov/abs/1206.7034}{{\tt
  arXiv:1206.7034}}].

\end{thebibliography}\endgroup

\end{document}